# Comparing local search paths with global search paths on protein residue networks: allosteric communication

Susan Khor
slc.khor@gmail.com
(July 11, 2016)

Although proteins have been recognized as small-world networks and their small-world network properties of clustering and short paths have been exploited computationally to produce biologically relevant information, they have not been truly explored as such, i.e. as navigable small-world networks in the original spirit of Milgram's work. This research seeks to fill this gap by exploring local search on a network representation of proteins and to probe the source of navigability in proteins. Previously, we confirmed that proteins are navigable small-world networks and observed that local search paths exhibit different characteristics from global search paths. In this paper, we investigate the biological relevance of the differences in path characteristics on a type III receptor tyrosine kinase (KIT). A chief difference that works in favour of local search paths as intra-protein communication pathways is their weaker proclivity, compared to global search paths, for traversing long-range edges. Long-range edges tend to be less stable and their inclusion tends to decrease the communication propensity of a path. The source of protein navigability is traced to clustering provided by short-range edges. The majority of a protein's short-range edges reside within structures deemed important for long-range energy transport and modulation of allosteric communication in proteins. Therefore, the disruption of intra-protein communication as a result of the destruction of these structures via random rewiring is expected. A local search perspective leads us to this expected conclusion while a global search perspective does not. These findings initiate the compilation of a list of path properties that are characteristic of intra-protein pathways and could suggest fresh avenues for evolving and regulating navigable (small-world) networks.

## 1. Introduction

Milgram tested the common intuitive idea of a small world with his social contact experiment [1]. The result of his experiment is remarkable because it revealed the power of a social network to create short paths covering large geographical distances between seemingly unrelated persons using only local information to build each link of a path. Watt's small-world network beta model with its random links [2] on the other hand is not navigable and various attempts have been made to evolve navigable small-world networks since, e.g. [3-6]. One early attempt to understand navigability in small-world networks was made by Kleinberg [7, 8]. The local search algorithm used in our work, which is called Euclidean distance Directed Search or *EDS* (section 2.5) is very much Kleinberg's algorithm (Milgram's protocol formalized) reinterpreted for our network representation of proteins, which is called Protein Residue Network or *PRN* (section 2.4).

In our previous paper [9], we found that PRNs are indeed navigable small-world networks. On a set of 166 PRNs built from PDB coordinate files [10], the average EDS path length of a PRN scaled logarithmically with PRN size. PRN size is the number of nodes in a PRN and a PRN node represents a protein residue or amino acid. Further, EDS paths differ significantly from paths found using breadth-first-search (*BFS*) in several ways. On average, EDS paths are more varied in length, are less diffusive (have lower search cost) and are less inclined to traverse long-range edges. The endpoints of a long-range PRN edge are close in 3D space but far apart (> 10 positions) on the protein sequence [11].

In this paper, we propose that these differences in path characteristics work in favour of EDS paths as a phenomenological model of intra-protein communication. First, we conduct an investigation into the differences between EDS and BFS path characteristics and elaborate on their respective abilities to capture key allosteric communication in a type III receptor tyrosine kinase (RTK), KIT. Second, we test the dependency of both EDS and BFS paths on the presence of structures that are recognized to play leading roles in either long-range energy transport or regulation of intra-protein communication (section 3.7).

## 2. Materials and Method

### 2.1 Allosteric communication in KIT

KIT is a kinase protein. Kinase proteins play a major role in signal transduction by turning cellular signaling pathways on and off through phosphorylation of substrate proteins. Some kinases are self-activating. For KIT, this happens when its extra-cellular ligand-binding domains bind with stem cell factor (SCF) [12] which results in phosphorylation of its tyrosine residues: Y568 and Y570 [13]. KIT can also be activated through point mutations, and such abnormal regulation of KIT has been implicated in several cancers.

The key allosteric characteristics of KIT that we will be examining are primarily based on the study by Laine et al. [14]. In that study, allosteric communication in the cytoplasmic region of KIT (PDB: 1T45, chain A, residues 547…694, 753…935) was derived from MD simulations and represented as a network of *independent dynamic segments* (section 2.2) connected by *communication pathways* (section 2.3). This network is sensitive enough to detect differences between KIT in its wild-type (WT) form which is autoinhibited and inactive, and the active form of KIT induced by an oncogenic point mutation, D816V (Aspartate at position 816 is replaced with Valine). In WT, the juxtamembrane region (JMR, residues 547…581) communicates with the spatially distant activation loop region (A-loop, residues 810…835) via the catalytic-loop (residues 790…797). This communication is interrupted when WT is activated by D816V, which triggers structural reorganization of the JMR and A-loop regions, and decreases communication between the N- and C- lobes of the protein kinase domain (PTK) of KIT. This change in communication pattern reflects the repositioning of the JMR in relation to the PTK, and the dimerization of the PTK upon KIT activation [13]. Communication between the JMR and the A-loop via the catalytic-loop is restored with a second point mutation, D792E (Aspartate at position 792 is replaced with Glutamate), which reestablishes a bond that Laine et al. [14] proposes is key to the inter-conversion of KIT between its active and inactive states: the H(hydrogen)-bond between residues Y823 and D792. Fig. 1 illustrates these mutation induced state changes in KIT.

Since there are no crystallized or NMR versions of the D816V mutant (MU) or the D816V/D792E double mutant (dbMU) to date, we rely on theoretical models of MU and dbMU (Appendix A). Throughout the paper, we use the terms 'MU' and 'dbMU' when making reference to the structures and findings from Laine et al. [13-15]; and use 1T45_MU and 1T45_dbMU when referencing our theoretical models and findings.

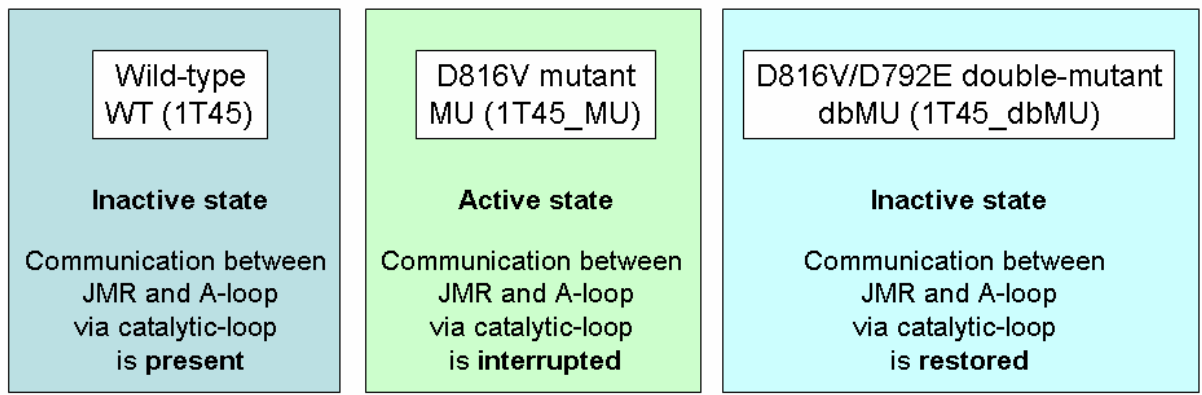

**Fig. 1** Activation and deactivation of KIT tyrosine kinase by point mutations D816V and D792E.

## 2.2 Independent dynamic segment (IDS)

An IDS is a set of residues close in 3D space, whose carbon-alpha (Cα) motions are highly correlated with each other but are largely decoupled from the rest of the system [14]. IDSs are grown from seed residues that are identified through Principal Component Analysis and Local Feature Analysis [16, 17]. IDSs correspond to well known functional regions distributed throughout KIT [14]. Residues of the 10 IDS for the wild type (WT), the D816V mutant (MU) and the D816V/D792E double mutant (dbMU) are available in Table S1 of [14] (Table H1). Fig. 2 shows where the WT IDSs, the JMR, the A-loop and the N- and C- lobes are located within a cartoon of 1T45.

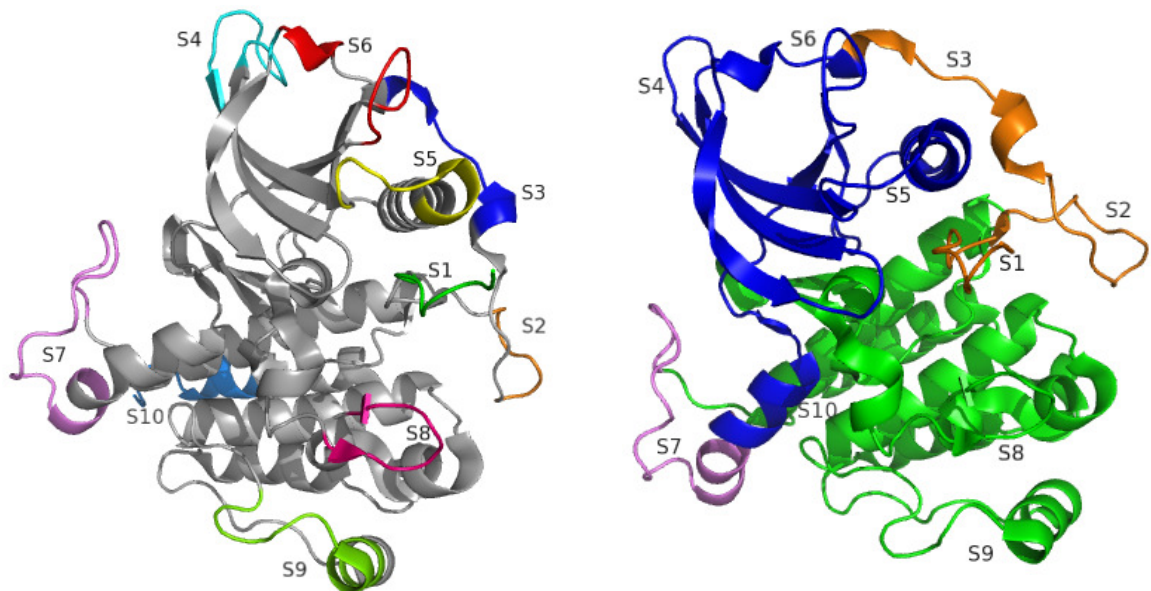

**Fig. 2** Left: Residues of the ten WT IDSs are colored differently on the cartoon of 1T45 and labeled S1, S2,…,S10. S1, S2 and S3 are in the JMR, and S8 is in the A-loop region. Right: A cartoon of 1T45 with the JMR colored orange, the N-lobe (582…684) blue, the C-lobe (762…935) green, and the pseudo-KID (685…694, 753…761) violet. S4, S5 and S6 are located in the N-lobe, while S8, S9 and S10 are in the C-lobe.

## 2.3 Communication pathway (CP)

Communication pathways are built upon the notion that the communication abilities of a protein's residues correlate with their motions at equilibrium [18]. A CP is composed of a chain of residues non-adjacently located on the protein sequence, such that each link in the chain is non-covalently bonded, stable, and the commute time between *any* pair of residues in the chain is small [14, 15]. A link is *stable* if it has high occupancy, i.e. is present in a large fraction (above a threshold e.g. $\geq 50\%$) of the protein's native ensemble (converged MD simulation of the protein's native dynamics). The *commute time* between a pair of residues $(i, j)$ is the variance of the Euclidean distance between the Cα atoms of $i$, and $j$ in the protein's native ensemble [14, 18]. A larger variance increases commute time for a residue pair and decreases its *communication propensity*.

## 2.4 Protein Residue Network (PRN)

A PRN is constructed from the coordinates obtained from either the PDB [10] or snapshots of a Molecular Dynamics (MD) run. A PRN is a simple undirected connected graph. Each node represents an amino acid molecule (residue) in a protein sequence. Let the number of nodes be $N$. Nodes are labeled by the residue id (*rid*) given in the coordinates file so that gaps in node labels correspond in size to gaps in sequence locations. In situations where it makes more sense to refer to a protein's residue than to its PRN node, the residue's amino acid single-letter code is prefixed to its rid, e.g. V560 is the Valine amino acid at protein sequence position 560.

Our PRN construction is based on the method in [19] which highlights the role side-chain atoms play in identifying well-formed protein structures. Two nodes $u$ and $v$ are linked if and only if $|u - v| > 1$ (peptide bonds are ignored), and their interaction strength $I_{uv}$ is above a threshold. $I_{uv} = \dfrac{n_{uv} \times 100}{\sqrt{R_u \times R_v}}$ where $n_{uv}$ is the number of distinct pairs $(i, j)$ such that $i$ is an atom of residue $u$, $j$ is an atom of residue $v$, and the Euclidean distance between atoms $i$ and $j$ is within a cutoff distance. $R_u$ and $R_v$ are extracted from a table of normalization values by residue type (Table 1 in [20]). When computing $I_{uv}$ to demonstrate the role side-chains play in specifying protein structure, only side-chain atoms were used in [19]. In contrast, we use both the side-chain and the protein backbone atoms of an amino acid since the protein backbone plays a significant role in allosteric communication [21-23]. Our cutoff distance is 7.5 Å and $I_{uv} \geq 5.0$. Values for these parameters were set through trial and error previously in [9], with the goal of creating PRNs that are singly connected without being unnecessarily dense. Ignoring peptide bonds is appropriate since validation of our model relies on results from [14, 15].

When necessary to distinguish PRNs by source, let *PRN0* be the PRN that is constructed from a protein's PDB file as opposed to a MD snapshot. The set of PRN edges is partitioned into short-range (*SE*) and long-range (*LE*) edges. An edge $(u, v)$ is long-range if and only if $|u - v| > 10$, and short-range otherwise [11].

A common network representation of a protein is a residue interaction network (RIN) in which a pair of nodes (each representing an amino acid) is connected by an edge if the Euclidean distance between its endpoint Cα atoms is within a user specified threshold range. Appendix G discusses the shortcomings of RINs compared to PRNs.

## 2.5 The Euclidean Directed Search (EDS) algorithm

EDS is a greedy local search algorithm with backtracking similar in principal to Kleinberg's [7]. Pseudo-code for the EDS algorithm is in Appendix B. At each step of a search, EDS surveys the proximity to target of the current node's direct neighbors in a PRN, and moves to a node $x$ not yet on the path which is closest (Euclidean distance) amongst all nodes surveyed so far to the target node. It is possible that $x$ is not a direct neighbor of the current node (it is a direct neighbor of some other node already on the path). In this case, EDS retraces its steps (*backtrack*) until $x$ becomes reachable.

Fig. 3 shows an EDS path (solid lines) connecting 559 and 823 in the WT PRN. A vastly reduced subset of direct neighbors is shown for each node; however if there is a PRN edge between nodes in Fig. 3, it is shown as a dashed line. The number of direct neighbors (degree) in the WT PRN for each node $x$ in Fig. 3 and the Euclidean distance (up to four decimal places, units in Angström Å) between $x$'s Cα atom and that of the target node 823, are listed on the right of Fig. 3. Starting at node 559, EDS moves to 557 because of all 559's direct neighbors, 557 is closest to 823. From node 557, EDS checks the direct neighborhood of both 559 and 557 to find an as yet unvisited node that is closest to 823, and moves to 792. This penultimate move creates the EDS path ⟨559, 557, 792, 823⟩.

BFS and EDS are run for all node pairs $(u, v)$ where $u \neq v$. The total number of paths for each is $N(N-1)$. The length of a path is the number of edges in it. Our C/C++ implementation, which is not optimized, took 205 (BFS) and 31 (EDS) seconds on a 32-bit personal computer running Windows OS to generate all EDS paths and all BFS paths for the 1T45 PRN, which has 331 nodes and 2314 edges. These execution times include time to output the paths to file. The EDS and BFS algorithms are used in their rudimentary form so that the natural character of their paths are revealed and compared. This avoids the complications arising from a method that uses a global search algorithm on a network that is embellished with additional information such that the paths produced are more similar in characteristics to local than global search paths.

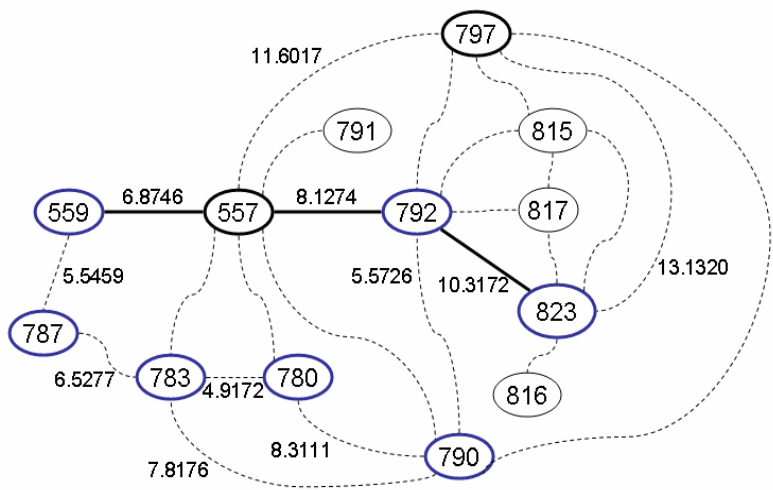

| Node | Degree | Cα-Cα Euclidean distance (Å) to 823 |
|------|--------|-------------------------------------|
| 557  | 22     | 16.9827 |
| 559  | 16     | 22.9016 |
| 780  | 24     | 21.6446 |
| 783  | 20     | 23.2078 |
| 787  | 15     | 25.3612 |
| 790  | 18     | 15.7188 |
| 791  | 16     | 12.8779 |
| 792  | 19     | 10.3172 |
| 797  | 20     | 13.1320 |
| 815  | 23     | 11.0238 |
| 816  | 11     | 10.1687 |
| 817  | 17     | 6.5876  |
| 823  | 20     | 0.0000  |

**Fig. 3** ⟨559, 557, 792, 823⟩ is an EDS path in WT that starts in the JMR, visits a node in the catalytic-loop (792) and terminates at a node in the A-loop (823). The length of this EDS path is three. The corresponding BFS paths are ⟨559, 557, 792, 823⟩ and ⟨823, 797, 557, 559⟩. The CP that connects the JMR and A-loop regions via the catalytic-loop in inactive KIT is ⟨559, 787, 783, 780, 790, 792, 823⟩ [15]. We revisit this CP and the EDS and BFS paths in section 3.3. The real numbers beside several of the edges give the Cα-Cα Euclidean distance between the endpoints of the edge.

There is a major difference between CPs and both EDS and BFS paths. CPs are akin to a constrained diffusion process from a source node (paths or chains linking residues are extended out from a node until there are no more edges with acceptable stability and commute time), than to an unconstrained targeted search for a node. Further, depending on the cutoff values used for required stability and acceptable commute time, there may be no CPs extending out from some residues. Ref. [15] reports that a maximum of 67 CPs extend from a WT residue, and the number of CPs for WT and MU (D816V KIT mutant) is 2404 and 1730 respectively. In contrast, EDS and BFS paths are possible, and algorithmically generated, for all pairs of nodes in a PRN.

Paths are partitioned into two sets by the sequence distance of their source and target nodes. Short-range paths (*SP*) are paths connecting source and target node-pairs within (≤) 10 residues apart on the protein sequence. Paths that are not short-range are long-range paths (*LP*).

## 2.6 Path stability and (estimated) path commute time

The definition of CPs in section 2.3 gives us two properties that an intra-protein communication path should likely possess: stability and communication propensity. Thus we need a way to quantify stability and commute time for an EDS or BFS path.

Per section 2.3, let $sb(e)$ be stability of a edge $e$, i.e. fraction of time it is present in a sequence of MD snapshots. Assuming edges of a PRN are independent of each other (this is not entirely true because of geometric constraints), we define *stability of a path p* with $n$ edges as $sb(p) = \prod_{i=1}^{n} sb(e_i)$. Paths with larger $sb(p)$ are more stable since all of its edges have high stability.

We define *path commute time* as the average commute time between *all* pairs of nodes on the path. This definition relies on the existence of edge commute time defined in section 2.3. When MD simulation data is not available, we use *estimated path commute time*, which for a path $p$ is the average Euclidean distance between *all* pairs of nodes on $p$. Estimated path commute time is based on the Cα-Cα Euclidean distance of residue pairs in a static protein structure, whereas path commute time is based on the commute time of residue pairs calculated from MD snapshots.

A path of length $\lambda$ has $\lambda(\lambda+1)/2$ node pairs. Some of the node pairs on a backtracking EDS path may not be distinct from each other, and commute time between a residue and itself or a sequence adjacent residue is zero. As an example, the estimated path commute time for ⟨783, 780, 790⟩ using the distance values in Fig. 3 is (4.9172 + 8.3111 + 7.8176)/3 = 7.0153.

## 2.7 Statistics and test of significance

Statistics (means, standard deviations, correlations) are produced with the R statistical package. Hypotheses are tested with R's one-sided t.test or wilcox.test command, paired when possible, and a p-value of < 0.01 is taken to confirm a hypothesis, unless stated otherwise.

## 3. Results and Discussion

### 3.1 Path stability and communication propensity.

Path stability and path commute time of EDS and BFS paths on the PRN0s of 12 randomly selected proteins from the Dynameomics database [24, 25] were evaluated (Appendix C). The emphasis here is on LP (long-range paths). Trivial paths, i.e. those directly connected to each other by a PRN edge, are excluded from this evaluation. Stability and commute time of edges were computed with snapshots from their native dynamics (298K) MD simulation per section 2.3, and these edge values were then used to calculate path stability and path commute time per section 2.6.

The EDS paths are significantly more stable and have significantly smaller commute times (i.e. better communication propensity) than BFS paths. This conclusion also holds when the analysis is broken down by path range. Both short- and long-range EDS paths are significantly more stable and have significantly better communication propensity than BFS paths of the same range (Fig. C1 & Table C2). Since having good stability and smaller commute times are two foundational characteristics of CPs, these findings support the notion that EDS paths are more plausible intra-protein communication pathways than BFS paths. SP (short-range paths) exhibit significantly better stability and significantly higher communication propensity than LP, for both EDS and BFS.

The findings above follow from differences in edge usage pattern previously observed in [9], and differences in stability and commute times of edges by different ranges (Table C2). Compared with BFS paths, EDS paths are less inclined to use LE (long-range edges) than SE (short-range edges). SE are significantly more stable and have significantly smaller commute times than LE. A pair of residues with a small commute time means their Cα-Cα Euclidean distance has not varied much over the MD simulation. It stands to reason that if a link exists between such a pair, the link is expected to be highly stable.

For PRNs without available MD simulation data to compute commute time of edges, we proposed in section 2.6 the use of estimated path commute time. To check that this is a reasonable thing to do, the following correlations, which are all significant (p-value < 0.01), were calculated with the 2EZN data: (i) Cα-Cα Euclidean distance of PRN0 edges and commute time of 2EZN edges is positively (0.2403457) correlated; (ii) path commute time is positively correlated with estimated path commute time (0.4241025 for EDS, 0.3975642 for BFS); (iii) path stability is negatively correlated with path commute time (-0.3500514 for EDS, -0.3670511 for BFS); and (iv) path stability is negatively correlated with estimated path commute time (-0.5074292 for EDS, -0.5451615 for BFS).

### 3.2 Path compressibility and path set reducibility.

The combination of IDS and CP capture two complementary ways a perturbation may propagate within a protein: via concerted local atomic fluctuations for short-range communication and via a network of well-defined pathways for long-range communication. To discover which path type better reflects this view of intra-protein communication as a network of modules connected in a well-organized manner, the *compressibility* and *reducibility* of both EDS and BFS paths are evaluated relative to random walks (RW) on the three KIT PRNs.

IDS nodes are relabeled to calculate the *compression ratio* (*cr*) of a path/walk (Appendix D). *cr* values range from 0.0 when there is no compression to 1.0 when there is maximum compression, i.e. a path/walk stays within a single IDS and is thereby compressed to a single node. On average, all three path types (RW, EDS and BFS) exhibit some level of path compressibility on an all three KIT PRNs (Fig. D1). That the RW paths are compressible supports the presence of modularity in the PRNs in the form of IDSs. In all three KIT PRNs, EDS paths are significantly more compressible than BFS paths. For each KIT PRN, EDS produces at least twice as many highly ($cr \geq 0.5$) compressed paths than BFS, and about 10 times more highly compressed paths than RW (Fig. D2). These observations imply that the modular parts of the PRNs in terms of their respective IDSs are better manifest by EDS paths than BFS paths. The 1T45_MU paths are significantly more compressible than paths from the other two KIT PRNs because MU has more IDS nodes.

In a well-structured communication network, there should also be well-organized inter-module routes. Reducibility indicates the level of organization in a set of paths and is quantified by the *reducibility ratio* (Appendix D). A better organized set of paths is more reducible since a larger proportion of its paths will be sub-paths of other paths in the set. In all three KIT PRNs, EDS paths are about 1.5 times more reducible than BFS paths (Fig. D3). The reducibility of BFS paths may be increased by reusing parts of already discovered shortest paths as a sub-path of a shortest path is also a shortest path. Nonetheless, a choice still needs to be made with regards to the shortest paths reused, even if it is random. This decision need not be taken with EDS as it naturally creates a unique path between any ordered pair of nodes. The compressibility and reducibility results provide evidence that EDS carves a more well-organized communication network on the KIT PRNs than BFS.

### 3.3 Communication between JMR and A-loop through the catalytic-loop

In WT, the JMR and the A-loop regions are allosterically connected by a CP which is disrupted by D816V and restored by D792E (Fig. 1). This CP, referred to as CP_WT, comprises seven residues: V559, C787, L783, M780, H790, D792 and Y823 [15] (Fig. 4). It is possible to trace the CP_WT path along the edges of each of the three KIT PRNs. The estimated path commute times for ⟨559, 787, 783, 780, 790, 792, 823⟩ in the three KIT PRNs are 12.5562 (WT), 12.5745 (1T45_MU) and 12.5061 (1T45_dbMU).

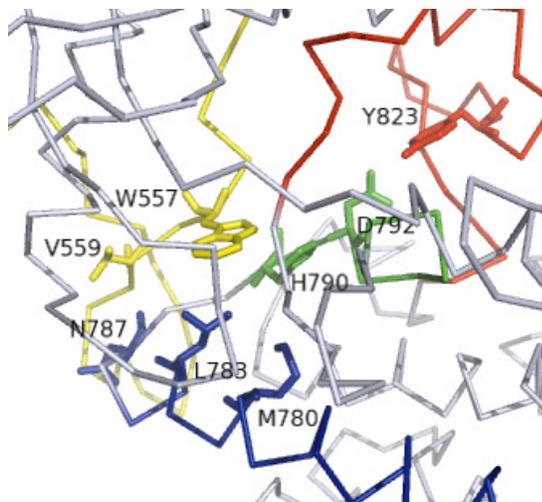

**Fig. 4** Residues involved in CP_WT and the JCA/ACJ paths that connects 559 with 823 are shown as sticks in a ribbon view of 1T45 (WT). The JMR, catalytic-loop, A-loop and E-helix are colored yellow, green, red and blue respectively.

Neither EDS nor BFS generated a path that traverses the CP_WT route. But both EDS and BFS did generate paths that connect the JMR and the A-loop through the catalytic-loop exclusively. We refer to such paths as JCA/ACJ paths. More formally, let $j_x$ be a residue in the JMR, $c_x$ be a catalytic-loop residue, $a_x$ be a residue in the A-loop region and $1 \leq p < q < k$, then a JCA path has the form ⟨$j_1$,…, $j_p$, $c_{p+1}$,…,$c_q$, $a_{q+1}$,…,$a_k$⟩ while a ACJ path takes the form ⟨$a_1$,…, $a_p$, $c_{p+1}$,…,$c_q$, $j_{q+1}$,…,$j_k$⟩. Since CP_WT also involves the E-helix (764…785), we could have defined JECA/ACEJ paths, where E stands for any E-helix residue, in a similar manner to JCA/ACJ paths. However, the set of JECA/ACEJ paths is empty for both EDS and BFS.

In all three KIT PRNs, almost all JCA/ACJ paths visited only one catalytic-loop node and the three most visited catalytic-loop nodes are 791, 792 and 797. Nodes 792 and 797 are part of the critical H-bond network mentioned in [14] (Table A3). Over half (52.38%) of WT's EDS JCA/ACJ paths but only a third (29.17%) of 1T45_MU's EDS JCA/ACJ paths pass through node 792. Node 791 is most popular with the EDS JCA/ACJ paths of both 1T45_MU (66.67%) and 1T45_dbMU (76.89%).

In all three KIT PRNs, node 559 is connected to node 823 by the same EDS JCA/ACJ path: ⟨559, 557, 792, 823⟩. In agreement with expected behavior, the estimated path commute time of this path is longer in the 1T45_MU PRN (13.2921) than in the WT (13.1855) or the 1T45_dbMU (13.1459) PRNs. In all three KIT PRNs, the BFS JCA/ACJ paths between nodes 559 and 823 are either ⟨559, 557, 797, 823⟩ or ⟨823, 792, 557, 559⟩. The average estimated path commute times for these paths are 14.04235 (WT), 14.2011 (1T45_MU) and 14.0328 (1T45_dbMU). Thus, the BFS JAC/ACJ paths between nodes 559 and 823 also exhibit the expected pattern of change in estimated path commute time.

Another JAC/ACJ path that shows signs of being disturbed in 1T45_MU and restored in 1T45_dbMU connects nodes 559 with 817. Node 817 is remarkable because it is adjacent to the point mutation site (D816V) and is part of the small-helix (817…819) that unfolds as a result of the point mutation. In both WT and 1T45_dbMU PRNs, the connecting EDS JCA path is ⟨559, 557, 792, 817⟩ with estimated path commute times of 13.4932 and 13.4274 respectively. In the 1T45_MU PRN, this path is ⟨559, 557, 792, 815, 817⟩ with an estimated path commute time of 13.8322. The addition of 815 is a consequence of there being no edges in the 1T45_MU PRN between nodes of the unfolded small-helix and any of the catalytic-loop nodes.

The above are two select examples of EDS JCA/ACJ paths that undergo changes in the different KIT PRNs. They echo the effects of the point mutations in MU and dbMU. The first example is inspired by CP_WT, and the second is newly observed by us. These communication effects are also detectable statistically at an aggregate level (Table 1). The estimated path commute time for EDS JCA/ACJ paths increased significantly in 1T45_MU relative to WT, and decreased significantly in 1T45_dbMU relative to 1T45_MU. The improvement in 1T45_dbMU is large enough that there is no significant difference between the estimated path commute time of WT and 1T45_dbMU EDS JCA/ACJ paths. In contrast, there is no significant difference in estimated path commute time between the WT, 1T45_MU and 1T45_dbMU BFS JCA/ACJ paths.

Table 1 The mean ± std. dev. estimated path commute time and SE fraction of EDS and BFS JCA/ACJ paths. SE fraction of a path is the fraction of edges on the path which is short-range.

| PRN | EDS | | BFS | |
| --- | --- | --- | --- | --- |
| | Est. path commute time | SE fraction of path | Est. path commute time | SE fraction of path |
| 1T45 (WT) | 13.30 ± 1.9748 | 0.4546 ± 0.2156 | 13.93 ± 2.0980 | 0.3001 ± 0.1710 |
| 1T45_MU | 13.74 ± 1.9696 | 0.4059 ± 0.2192 | 13.79 ± 2.1636 | 0.2850 ± 0.1975 |
| 1T45_dbMU | 13.20 ± 1.9995 | 0.3745 ± 0.2171 | 13.87 ± 2.0841 | 0.2580 ± 0.1893 |

In all three KIT PRNs, compared with the BFS JCA/ACJ paths, a significantly larger fraction of the EDS JCA/ACJ paths are composed of SE (Table 1). Having shorter estimated path commute time and utilizing more SE bodes well for both path stability and path commute time. SE tend to be more stable than LE, and path stability is negatively correlated with both path commute time and estimated path commute time (section 3.1). Findings in this section provide evidence that disruption in communication between the JMR and A-loop region through the catalytic-loop as a result of D816V and its subsequent restoration by D792E are better captured by EDS than BFS.

### 3.4 Identifying hub residues

Ref. [14] defines *hub* residues as residues that lay on the intersection of many CP and identified 71 such hub residues (Fig. 5-left). These hub residues are either evolutionarily conserved or have been observed to participate in the regulation of other receptor tyrosine kinases and cytoplasmic kinases. In complex network science, a hub node refers to a node with a large number of edges incident on it. The 71 hub residues are also hub nodes in this conventional sense (section 3.6).

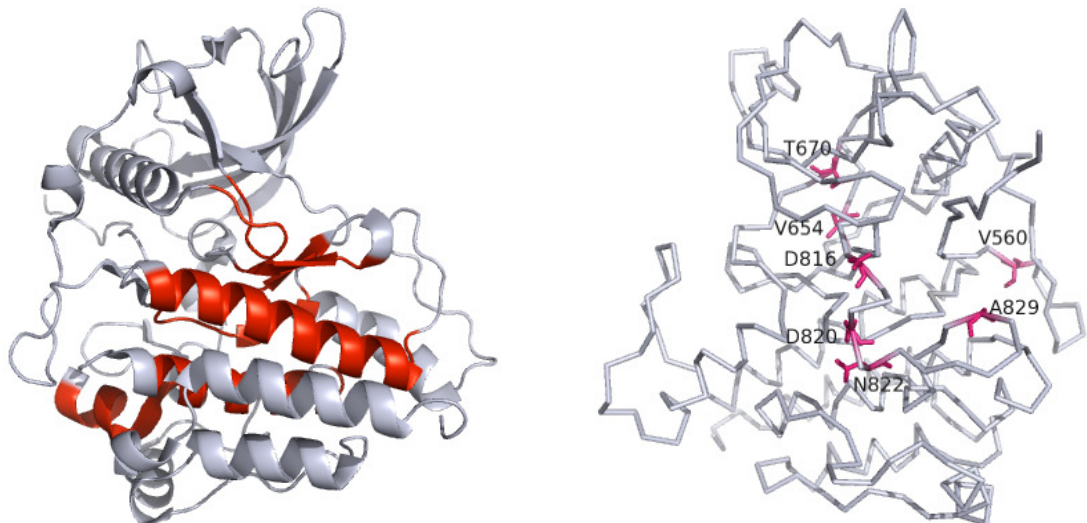

**Fig. 5** Left: The 71 hub residues identified in [14] are in red on a cartoon of 1T45. They are: 649…655 (C-loop-2), 764…785 (E-helix), 790…797 (catalytic-loop), 804…808 (β-strand B8), 835…843 (P+1 loop), 850…865 (F-helix), and 678, 798, 799, 800, 858, and 862 (catalytic spine). Right: The seven mutational hotspots referenced in [14] are shown as magenta sticks on a ribbon representation of 1T45.

The task here is to identify the 71 hub residues using information generated by EDS and BFS on the 1T45 PRN. A number of complex network approaches have been utilized to identify key residues for different purposes in a network of interacting protein residues. These approaches typically employ some notion of network centrality either by node degree, number of paths (betweenness), graph distance (closeness) or some combination of these [26-28]. We tried several of these strategies and found closeness based on estimated path commute time performed the best. This follows, since hub residues display fast commute times [14].

*Closeness* for a node $x$ is the total estimated path commute time from $x$ to all other nodes in the network, and from all other nodes in the network to $x$. $Closeness(x) = \sum_{y \neq x}^{N}[epct(x, y) + epct(y, x)]$. It is not necessarily the case that $epct(x, y) = epct(y, x)$ since a BFS path from node $v$ to node $w$ may differ from a BFS path from node $w$ to node $v$, and the EDS path from node $i$ to node $j$ may differ from the EDS path from node $j$ to node $i$. Nodes with smaller closeness values are nearer on average to all other nodes in a network. To check the reasonableness of this metric, we computed closeness for residues within the two lobes of WT's PTK and found as expected, that residues residing in the same lobe are closer to each other than residues residing in different lobes (Appendix E).

To recover the hub residues, nodes are sorted in non-decreasing order of their closeness values. We find that nodes with smaller closeness values are enriched with hub residues. Compared with all nodes, the hub nodes have significantly smaller closeness values while the (WT) IDS nodes have significantly larger closeness values (Fig. 6). It is possible to recover at least 60% of the 71 hubs residues at the cost of 26 false positives (10% of 260) with both EDS and BFS closeness values (Fig. 6). Full recovery is slightly quicker with EDS closeness (61.15% FPR) than with BFS closeness (66.92%). The eight hub residues in the catalytic-loop are amongst the easiest to recover. They reside in the top 13 nodes with the smallest EDS closeness value, and in the top 18 with the smallest BFS closeness value. In short, EDS performs as well as, if not slightly better, than BFS on this task.

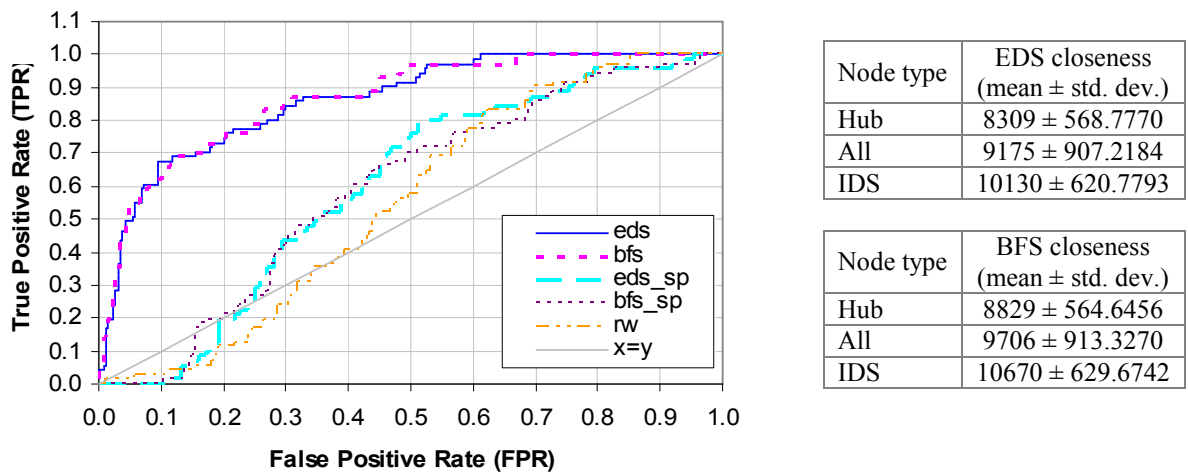

**Fig. 6** Left: Recovery of the 71 hub residues using closeness based on estimated path commute time. Right: EDS and BFS closeness summary statistics for node by type. Hub nodes are PRN nodes that represent hub residues; IDS nodes are PRN nodes that represent IDS residues.

Both EDS and BFS closeness outperformed RW closeness convincingly. Using closeness based only on SP (short-range paths) to recover hub residues is not a good strategy as evidenced by the poor performance of both *eds_sp* and *bfs_sp*. This is as expected since CP are channels for long-range communication.

### 3.5 Identifying mutational hotspots

Apart from D816, several other gain-of-function point mutations have been documented to induce irregular state change in KIT. These mutational hotspots include V560, V654, T670, D820, N822 and A829 [14] (Fig. 5-right). The adverse effects of these mutations can be suppressed by kinase inhibitors, e.g. Imatinib is effective against mutations at position 560, and sunitinib is effective against mutations at positions 654 and 670. However mutations at position 816 have so far proven resistant to both types of kinase inhibitors [14, 15].

We noticed from our attempts to recover hub residues in section 3.4 that residues located next to a mutational hotspot in the KIT protein sequence tend to have large *node betweenness*. Nodes with larger betweenness values are traversed by more paths. Therefore instead of trying to recover the seven mutational hotspots directly, we extended the target set to seven triples. Each triple comprises the hotspot residue and its immediate neighbors on the protein sequence. Our extended target set of residues is {(559, 560, 561), (653, 654, 655), (669, 670, 671), (815, 816, 817), (819, 820, 821), (821, 822, 823), (828, 829, 830)}. We need only recover one member from each triple to achieve 100% recovery.

Using this strategy of an extended target set and sorting the nodes in non-ascending order of their betweenness values, full recovery was possible with a False Positive Rate (FPR) of 26.54% (incurring the cost of 86 false positives). This best performance is yielded by EDS betweenness calculated using short-range paths only (*eds_sp* in Fig. 7). However, full recovery may not be necessary and the complete set of mutational hotspots may not be known *a priori*. At smaller FPRs, *eds_sp* also yielded the best performance. Small FPRs are preferably *per se*, but more so for this problem since there are potentially three tests associated with each candidate residue: the residue itself and its two sequence neighbors. This is where heuristics based on amino-acid chemistry and domain expertise can help further *in silico* winnowing of the candidates.

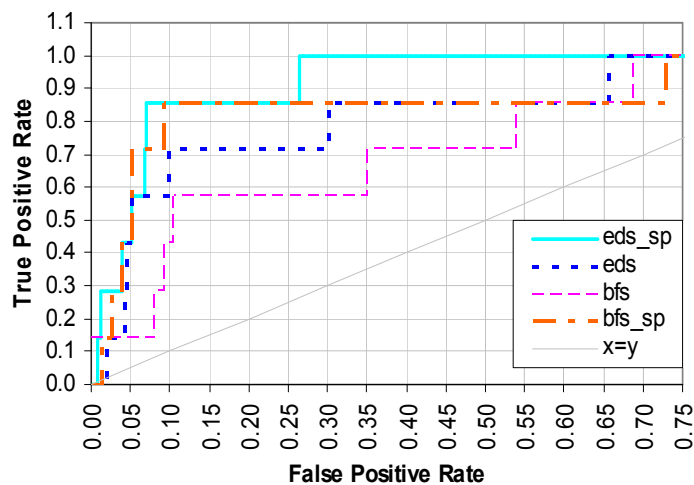

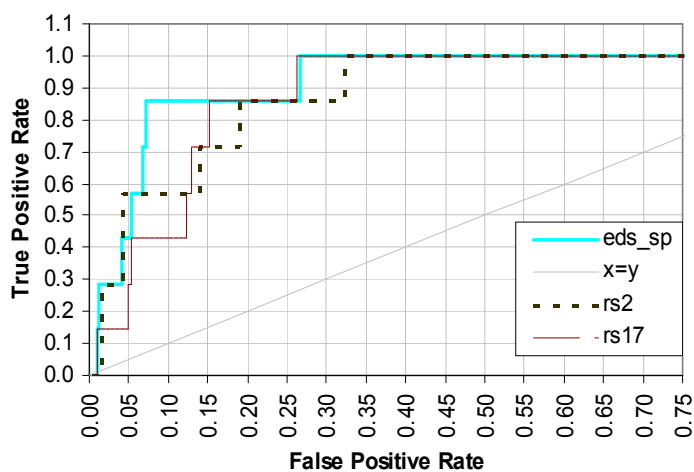

**Fig. 7** Top: Recovery of the seven mutational hotspots using node betweenness and an extended target set. Bottom: *eds_sp* outperforms random ordering of nodes. *rs2* and *rs17* are the two best results from 20 random node permutations.

There are also qualitative differences in the order in which the mutational hotspot residues are recovered. The target residues and their respective ranks are listed in Table 2. Residues that neighbor 816 to its immediate left and right on the protein sequence (815 and 817) are easier to recover when all paths are used to compute node betweenness (*eds* and *bfs* columns in Table 2). Residue 791, the left sequence neighbor of residue 792, which is mutated to restore communication interrupted by D816V, is also ranked highly by both *eds* (rank 5) and *bfs* (rank 12). In contrast, residue 560, which borders IDS S2 in the JM-Switch region, and whose mutation is studied extensively in [15], is recovered most easily with *eds_sp*. This difference in effectiveness of short- and long-range paths may be a reflection of the difference in impact range of the mutational hotspots. While mutations at 816 produced long-range structural effects [14, 15], mutations at 560 did not appear to have a structural effect on the distant A-loop [15].

For both EDS and BFS, node betweenness computed using short-range paths only (*eds_sp* and *bfs_sp* columns in Table 2) produced better results than node betweenness based on all paths (Fig. 7-top). We attribute this difference to the proximity of the mutational hotspots to IDSs. It may be that using only short-range paths to compute betweenness is helping to clarify the centrality of a node to the short-range communications around or within IDSs. Ref. [29] proposes that oncogenic mutation sites in protein kinases harness the local frustration present in the more flexible regions of proteins where the oncogenic mutations are commonly found, to produce global effects.

In contrast to the previous problem of recovering hub residues, the results produced with EDS betweenness for this task differs substantially from the results produced with BFS betweenness. We attribute this divergence to increased problem difficulty. Compared to the hub recovery problem, the mutational hotspot recovery problem has fewer targets (7 vs. 71) in the same search space and therefore any difference in rank becomes more noticeable. With fewer needles in the haystack, we were concerned about the competitiveness of a random strategy, i.e. ranking the residues in random order. 20 permutations were generated and we report the best two in Fig. 7-bottom. Permutation *rs2* performed well at low FPRs, but took a long time to fully recover all targets. On the other hand, permutation *rs17* was slow to start but finished as quickly as *eds_sp*. It appears then that *eds_sp* gives the best of both worlds and is thus a better strategy than random.

**Table 2** Extended set of target mutational hotspot residues and their respective centrality rank. A more central residue has a larger betweenness value and a smaller/higher rank. Each bolded residue number increases the True Positive count by one to a maximum of seven in Fig. 7.

| *eds* | | *eds_sp* | | *bfs* | | *bfs_sp* | |
|---|---|---|---|---|---|---|---|
| Rank | Residue | Rank | Residue | Rank | Residue | Rank | Residue |
| 8 | **817** | 4 | **560** | 1 | **815** | 6 | **669** |
| 9 | 815 | 6 | **669** | 3 | 817 | 7 | 670 |
| 16 | **669** | 16 | **653** | 28 | 823 | 11 | **559** |
| 18 | **653** | 21 | **815** | 33 | **671** | 16 | **655** |
| 21 | **823** | 22 | 670 | 37 | 670 | 21 | **823** |
| 25 | 671 | 26 | 817 | 38 | **655** | 22 | **830** |
| 35 | 654 | 27 | **823** | 40 | 654 | 36 | **817** |
| 37 | **830** | 29 | **830** | 49 | 653 | 40 | 560 |
| 59 | 670 | 43 | 655 | 53 | 669 | 97 | 561 |
| 64 | 655 | 93 | **820** | 64 | **822** | 101 | 815 |
| 77 | 822 | 125 | 559 | 119 | **560** | 134 | 653 |
| 104 | **560** | 130 | 816 | 127 | 559 | 135 | 671 |
| 117 | 559 | 156 | 671 | 181 | **830** | 152 | 654 |
| 181 | 828 | 230 | 829 | 230 | **821** | 208 | 816 |
| 203 | 829 | 249 | 561 | 235 | 816 | 243 | **820** |
| 216 | 816 | 253 | 828 | 264 | 820 | 288 | 821 |
| 220 | **820** | 295 | 822 | 283 | 561 | 303 | 829 |
| 261 | 821 | 308 | 821 | 294 | 829 | 311 | 822 |
| 296 | 561 | 315 | 819 | 317 | 819 | 327 | 819 |
| 321 | 819 | 322 | 654 | 329 | 828 | 328 | 828 |

**3.6 Attributes of structures mediating allosteric communication in KIT**

This section examines the secondary structure and network characteristics of IDS, hub and mutational hotspot residues to understand how their structural attributes correlate with the roles they play in intra-KIT communication.

The term 'independent dynamic segment' appeared previously in [30] within the context of protein binding sites and their ability to propagate signals within a protein a great distance away from the original point of perturbation. This ability was associated with the notion of 'discrete breathers' (DB) which are sites that can be found in a spatially discrete and topologically inhomogeneous network of oscillating nodes. Within the context of proteins, namely the nonlinear network model, DB sites were observed to harvest energy from the background of the system and other network sites (energy is redistributed in favour of DB sites) [31], and to 'jump' to other sites distantly located in the network with good energy yield and specificity under suitable conditions [32]. DB sites tend to form in the stiffest parts of a protein [31], and energy is transferred most efficiently along rigid sites [32].

Ref. [33] characterized the rigidity of a site in terms of its node degree and node clustering coefficient, and found that the energy gap (excitation threshold) of a DB is negatively correlated with node degree, and positively correlated with node clustering coefficient. The degree of a node is the number of edges incident on the node in a network, viz., a PRN which is a simple, undirected and unweighted graph. The clustering coefficient of a node gives the edge density of the node's direct neighborhood, e.g. the clustering coefficient for a node with degree $k$ is $2e/(k(k-1))$ where $e$ is the number of edges between the $k$ nodes. (An alternative way to quantify the level of clustering in a network is to count the number of triangles or cycles of length three in the network, which is done in [33].) DBs spontaneously emerge more easily at sites with lower excitation thresholds. In short, DB sites form more readily in the stiffer parts of a protein which tend to be richly connected globally but poorly connected locally, and less readily in the more flexible parts of a protein which tend to be poorly connected globally but richly connected locally.

The aforementioned characteristics of an 'independent dynamic segment' stand in stark contrast to the characteristics of an IDS, and more fittingly describe the hub residues instead. Each IDS has a solvent-exposed loop component which gives it the flexibility to have its own independent dynamics (Fig. 2). In contrast, the hub residues tend to be situated in the more rigid parts of a protein (Fig. 3). The IDS nodes are predominantly turn-residues, while the hub nodes are dominated by helix-residues (Table 3). This difference in rigidity profile between IDS and hub nodes for KIT was also observed in four other proteins [34]. Secondary structure is assigned by DSSP-2.0.4 [35, 36]. The DSSP assignments are partitioned into three basic groups. *Helix-residues* (H) are residues assigned either 'H', 'I' or 'G', *beta-residues* (B) are residues assigned either 'E' or 'B', and *turn-residues* (T) are neither helix- nor beta-residues, i.e. encompasses coils, loops, turns and other disordered regions.

**Table 3** Secondary structure characteristics of IDS, hub and mutational hotspot residues.

| KIT variant | IDS residues | | | Hub residues | | | Mutational hotspots | | |
|---|---|---|---|---|---|---|---|---|---|
| | H | B | T | H | B | T | H | B | T |
| 1T45 (WT) | 24 | 9 | 68 | 46 | 8 | 17 | 0 | 2 | 5 |
| 1T45_MU | 26 | 19 | 72 | | | | | | |
| 1T45_dbMU | 29 | 8 | 72 | | | | | | |

In accord with their positions on opposite ends of the rigidity scale, IDS nodes exhibit significantly smaller node degree and significantly stronger node clustering than hub nodes. Table 4 and Fig.8-left report on the network characteristics of the WT nodes. The node degrees of all three KIT PRNs are positively and strongly correlated with each other (> 0.97); and so are the node clustering coefficients (> 0.87). The IDS and hub nodes not only differ in the number of edges but also the composition of edges incident on them. A significantly larger fraction of edges incident on IDS nodes are SE (short-range edges), while a significantly larger fraction of edges incident on hub nodes are LE (long-range edges) (Table 5). In light of the discussion so far, both the IDS and hub residues are well-suited structurally for their respective roles in short- and long-range in intra-KIT communication.

**Table 4** Network characteristics of 1T45 (WT) PRN nodes (mean ± std. dev.)

| Node type | Number of nodes | Node degree | Node clustering coefficient |
|---|---|---|---|
| All | 331 | 13.98 ± 5.2766 | 0.3900 ± 0.1019 |
| Hubs | 71 | 18.01 ± 4.3408 | 0.3568 ± 0.0483 |
| IDS | 101 | 9.743 ± 4.1028 | 0.4402 ± 0.1388 |
| Mutational hotspots | 7 | 13.43 ± 4.8599 | 0.4045 ± 0.1139 |

**Table 5** Composition of edges incident on different node types in the 1T45 (WT) PRN (mean ± std. dev.)

| Node type | Fraction of node edges that are SE | Fraction of node edges that are LE |
|---|---|---|
| All | 0.5191 ± 0.2214 | 0.4809 ± 0.2214 |
| IDS | 0.6709 ± 0.2229 | 0.3291 ± 0.2229 |
| Hubs | 0.4350 ± 0.1490 | 0.5650 ± 0.1490 |

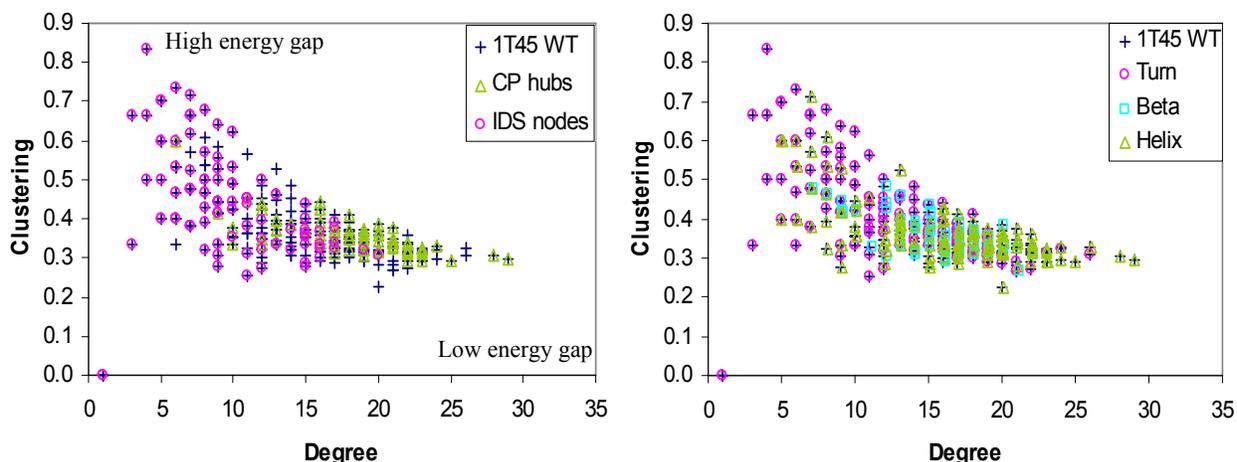

**Fig. 8** 1T45 residues by their node degree and node clustering coefficient in the WT PRN. Lower (higher) energy gap (DB excitation threshold) is associated with more (less) rigidity, which is indicated by larger (smaller) node degree and smaller (larger) node clustering coefficient [33].

The other aspect of allostery is the regulation of communication. It was previously observed that oncogenic mutations in protein kinases tend to appear in the more flexible or locally frustrated regions [29]. Five of KIT's mutational hotspots are turn-residues with the remaining two beta-residues (Table 3). The mutational hotspots do not have a significantly larger average node degree (Table 4), and only one of them (V654) is a hub residue. Further, the mutational hotspots are more central to short- than to long-range EDS paths (Fig. 7). These structural attributes of KIT's mutational hotspots reveal a more sutble tactic to regulate communication in a small-world network, i.e. obliquely through nodes that appear unassuming (are not central to long-range communication or have many direct neighbors) instead of a head-on hub-centric attack. Although less obvious, this tactic is understandable. The hubs are mainly located in the more rigid (helix) structures. While ridigity enhances communication efficiency and therefore range of a protein structure, it also increases its brittleness. Thus, to deform a protein without destroying its long-range communication network, which may still be needed to propagate the effects of the deformation to other parts of the protein or to other proteins, it is logical to capitalize on the more malleable parts of a protein, as the mutational hotspots do.

The structural analysis on IDS and hub nodes can be generalized to residues by secondary structure type. The WT helix-residues have a significantly larger node degree and a significantly smaller node clustering coefficient than the WT turn-residues (Fig. 8-right, Table G2). The WT helix-residues also have significantly smaller closeness values than the WT turn-residues. The energy transport role of helix structures in proteins (not necessarily through the DB mechanism) has been observed directly in experimental studies, e.g. [21, 22]. Ref [18] observed from the five enzymes examined that helix-residues tend to have better communication efficiency than turn-residues. But despite their general lower communication efficiencies and shorter communication range, the KIT case study reminds us that some turn-residues can be crucial to allosteric communication. Both the IDS and mutational hotspot residues are enriched with turn-residues (Table 3). The catalytic-loop (790-797) is central to allosteric communication in KIT. Notwithstanding its name, three (795…797) of its eight residues are helix-residues. The catalytic-loop residues have a significantly larger node degree and a significantly smaller node clustering coefficient, which places them on the stiffer end of the rigidity scale.

### 3.7 Appropriate path dependency on structures mediating communication

The discussion in section 3.6 points to the turn rich IDSs and the helix rich hub residues as protein structures noted for mediating intra-protein communication in KIT. Here, we would like to disrupt these communication mediating structures to investigate the dependency of EDS and BFS paths on the integrity of these structures.

To facilitate this investigation on a larger set of proteins, we create randSE and randLE networks by randomly rewiring the endpoints of SE (short-range edges) and LE (long-range edges) in PRNs such that the resulting network remains a simple graph, node degree is conserved and no edge connects residues consecutively located on the protein sequence[1]. From studying KIT's WT PRN edges, we observed that the SE cover 89.41% of all intra-IDS edges and 97.94% of all intra-SSE edges (Fig. 9). Of the short-range intra-SSE edges, 55.75% are within helices and 38.25 % are within turn SSEs. An intra-IDS edge is a PRN edge whose endpoints are nodes of the same IDS. An intra-SSE edge is an edge whose endpoints are within the same secondary structure element (SSE). SSEs locate alpha-helices, beta-strands and linear stretches of turn-residues in a protein sequence. SSEs are identified as follows: a SSE is a set of at least three residues, consecutively located on the protein sequence, which have the same (DSSP) secondary structure assignment [34]. From the composition of edge sets in Fig. 9, a randSE network is expected to be more disruptive to communication mediating structures than a randLE network, and accordingly a randSE network should produce significantly longer paths on average. The contact maps in Fig. F1 illustrate the edge shuffling effect on KIT's WT PRN.

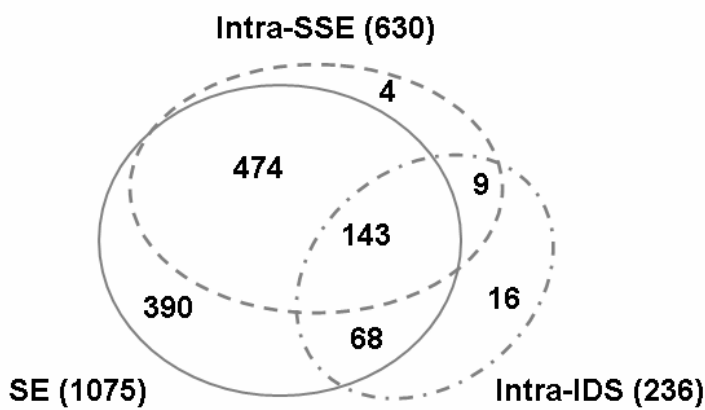

**Fig. 9** Relationship of KIT's short-range, intra-SSE and intra-IDS edges shown as a Venn diagram (not to scale).

KIT's WT EDS paths are significantly longer in its randSE network, but are not significantly longer in its randLE networks (Table 6). In contrast, KIT's WT BFS paths are significantly shorter in both its randSE and randLE networks. This means that the EDS paths exhibit appropriate dependency on KIT's communication mediating structures while the BFS paths do not. This conclusion holds for the 166 other proteins examined (Fig. F2 top left and right). The average lengths of EDS paths for this set of proteins increased to the extent that they no longer scale logarithmically with network size. As such randSE networks are not navigable. Interestingly, rewiring only the long-range edges of PRNs does not destroy their navigability. Like their PRNs, randLE networks are navigable.

**Table 6** EDS and BFS path length and estimated path commute time statistics.

| KIT (WT) | EDS path length | | EDS estimated path commute time | | BFS path length | | BFS estimated path commute time | |
|---|---|---|---|---|---|---|---|---|
| | Mean | Std. dev. | Mean | Std. dev. | Mean | Std. dev. | Mean | Std. dev. |
| PRN | 4.5578 | 2.0760 | 13.9008 | 4.0553 | 3.7077 | 1.3716 | 14.7060 | 4.0071 |
| randSE | 12.9480 | 63.8351 | 14.9411 | 5.8459 | 2.5811 | 0.6207 | 20.5246 | 7.2604 |
| randLE | 4.5774 | 2.6192 | 15.0843 | 5.7357 | 2.6228 | 0.6814 | 20.0722 | 6.6611 |

Estimated path commute time of both EDS and BFS paths increased significantly in both randSE and randLE networks. Therefore the disruptive effect of edge rewiring on paths could not be distinguished with this measure. Nonetheless, the significantly longer estimated path commute time of EDS paths on KIT's WT randLE network is interesting because it may help explain why when both are navigable, KIT's WT PRN and not its randLE configuration is the native one[2].

---

[1] Due to [11], we have used 10 as the cutoff for short-range edges and paths throughout our research. It would be interesting to see the sensitivity of our results to this cutoff value.

PRN average path length is dominated by the length of LP (long-range paths) since there are more LP than SP (short-range paths) in a PRN. When analyzed separately (Table 7), both short- and long-range EDS paths are still significantly longer in a randSE network. Thus randomizing SE affects short- and long-range EDS paths in this same way. Average EDS path length does not change significantly in a randLE network because although randomizing LE increases the length of short-range EDS paths significantly, it has no substantial effect on the length of long-range EDS paths.

**Table 7** Effects of rewiring PRN edges on the length of short-range (SP) and long-range (LP) EDS and BFS paths.

| KIT WT | EDS | | BFS | |
|--------|-----|-----|-----|-----|
|        | SP  | LP  | SP  | LP  |
| randSE | ↑   | ↑   | ↑   | ↓   |
| randLE | ↑   | ~   | ↓   | ↓   |

↑ Significant increase
↓ Significant decrease
~ No significant change

Randomizing SE affects BFS LP and SP differently. While long-range BFS paths are still significantly shorter in a randSE network, short-range BFS paths become significantly longer. This deviation is in accordance with [37], which views a small-world network as a structure charged with both local and global communication efficiencies. Local communication efficiency is a function of the average (BFS) path length of subgraphs centered on every node in a network. Since short (BFS) path lengths amongst the direct neighbors of a node $x$ correlates with a large clustering coefficient for $x$, higher levels of network clustering is associated with better local communication efficiency. Randomizing LE shortens both BFS LP and SP significantly.

## 3.8 PRN navigability

Current theory of navigability in a small-world network rests on the assumption that local (short-range) and global (long-range) connections are distinguishable from each other. Local connections link nodes that are more similar to each other in some sense. In many theoretical small-world network models, locality is measured as distance, e.g. Manhattan distance in Kleinberg's grid model [7]. And the same distance norm is used to direct local searches in the network. In PRNs, we use edge sequence distance as a measure of locality, but we use Euclidean distance to direct EDS. Hence some positive correlation between sequence distance and Euclidean distance of edges is assumed. The average Spearman correlation between edge Euclidean distance and edge sequence distance for the 166 PRNs in [9] and their randSE and randLE networks are 0.4075 (std. dev. 0.1163), 0.5129 (std. dev. 0.1143) and 0.8011 (std. dev. 0.0570) respectively. All correlation values are significant. In contrast, there is almost zero correlation between edge Euclidean distance and edge sequence distance in random geometric networks.

The *random* edge rewiring process affects the ratio of long- to short-range edges (|LE|/|SE|), edge sequence distance distribution and node clustering coefficient distribution. |LE|/|SE| is a significant metric as it influences the distribution of edge sequence distance. |LE|/|SE| is fairly constant with network size for the 166 PRNs and their randLE networks, but increases with network size in the randSE networks (Fig. F2 middle-left). Numerous studies have observed or subscribed to the notion that a right skewed edge length distribution, i.e. the probability of an edge is inversely related to its length, is necessary for navigability of small-world networks [7, 38]. We use the coefficient of variation COV = std. dev. / mean

---

[2] The coordinates of nodes are not changed by the rewiring; so strictly from this aspect PRN and randLE have the same configuration. The randLE results here suggest the existence of some (how much?) room for the nodes to wiggle without adversely affecting navigability.

of edge sequence distances as a proxy for skew-ness of the edge sequence distance distribution. Since median edge sequence distance is less than mean edge sequence distance for the 166 PRNs, a larger COV implies a skew to the right, or a longer right-tail. The edge sequence distance COVs for the 166 PRNs and their randLE networks range between 1.0 and 2.0, but they are below 1.0 for the randSE networks (Fig. F2 middle-right). KIT's WT randSE network has a larger |LE|/|SE| and a smaller edge sequence distance COV (Table 8), subsequently a less right skewed edge sequence distance distribution (Fig. 10 top-left).

**Table 8** Network structural characteristics of the WT PRN, an instance of its randSE network and an instance of its randLE network. COV = std. dev./mean is the coefficient of variance.

| KIT (WT) | |LE|/|SE| | Edge sequence distance | | | | Node clustering | | |
|---|---|---|---|---|---|---|---|---|
| | | Median | Mean | Std. dev. | COV | Mean | Std. dev. | Max |
| PRN | 1.1525 | 14.00 | 46.42 | 63.3847 | 1.3655 | 0.3900 | 0.1019 | 0.8333 |
| randSE | 8.8051 | 72.00 | 102.40 | 90.8341 | 0.8871 | 0.1109 | 0.0624 | 0.3333 |
| randLE | 1.0998 | 17.00 | 78.70 | 98.6237 | 1.2532 | 0.1676 | 0.1254 | 0.7000 |

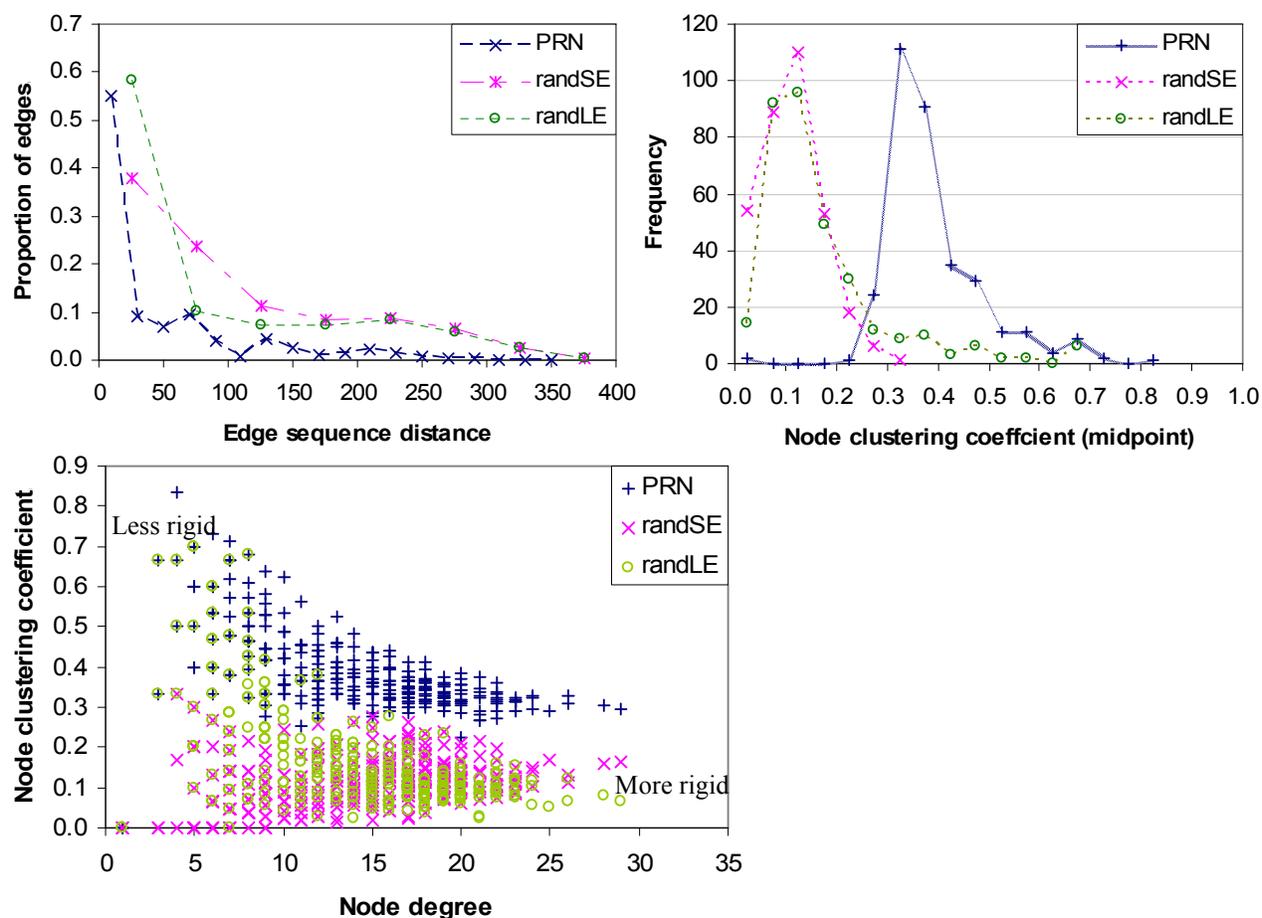

**Fig. 10** Top: Edge sequence distance distribution (left) and node clustering coefficient distribution (right) for the 1T45 (WT) PRN and its randSE and randLE networks. Bottom: Rigidity profiles of KIT's WT PRN, randSE and randLE networks measured by node degree and node clustering coefficient.

The clustering coefficients of the 166 PRNs are significantly reduced by edge rewiring (Fig. F2 bottom-left). However, maximum node clustering is significantly larger in the randLE networks which are navigable than in the randSE networks which are not navigable (Fig. F2 bottom-right). A wider node clustering coefficient distribution helps to promote diversity in terms of topological rigidity, and such diversity is a contributing factor in successful allosteric communication in protein kinases [29]. We have seen this principle demonstrated in KIT with its mix of flexible IDSs and rigid hub residues. Like the 166 PRNs, KIT's WT randLE network also has a larger maximum node clustering coefficient than its randSE network (Table 8), and consequently a wider node clustering coefficient distribution is possible for randLE nodes (Fig. 10 top-right). KIT's WT randLE (randSE) network has a broader (narrower) rigidity profile than randSE (randLE) (Fig. 10 bottom), and this difference in rigidity profile coincides with randLE's navigability (randSE's non-navigability). In contrast, observations with BFS average path length would lead one to conclude that a narrower rigidity profile is preferable for allosteric communication. A study of PRNs as a hybrid of SE and LE or rigid and flexible sub-networks may prove interesting.

### 3.9 Suggestions for future local search strategies on PRNs

The aim of this work is not to reproduce CPs or to replicate the findings in [13-15] with EDS paths on PRNs. Nonetheless, conducting this research has suggested additional search objectives that may improve the efficiency and/or the accuracy (e.g. paths that more closely resemble CPs) of local search on PRNs. Currently, the only objective of EDS is to minimize (in a greedy way) the Euclidean distance to target node, and backtracking is used as an escape from local minima.

The discussion on the communication adeptness of helix residues and their network structural characteristics (sections 3.6) suggests that a local search strategy on PRNs could benefit from node degree and/or secondary structure type information. The explicit use of node degree to guide local search has been demonstrated in other complex networks [38]. The influence of node degree on both EDS and BFS paths already happens implicitly, as evidenced by the presence of hierarchical EDS and BFS paths. A path with monotonically increasing, or monotonically decreasing, or monotonically increasing then monotonically decreasing node degrees is hierarchical [39]. For the 166 PRNs in [9], on average, 37.5% (std. dev. 12.0%) of all EDS and 50.7% (std. dev. 15.1%) of all BFS paths are hierarchical. These proportions may seem on the low side but keep in mind that proteins are biological and dynamic systems, so a less strict application of the definition of hierarchical paths could be more appropriate. A large (60%) proportion of the hierarchical EDS and BFS paths have a zoom-in zoom-out navigational pattern, i.e. their node degrees monotonically increase then monotonically decrease, which is a hallmark of successful local navigation in other real-world networks [40].

In terms of graph distance, EDS paths tend to be longer than BFS paths on average. But because EDS is less inclined to use long-range edges than BFS, EDS paths are more stable and have smaller estimated path commute time than BFS paths in general (section 3.1). From the perspective of wiring cost in a spatial network, long-range edges are seen as more expensive than short-range edges. Thus the addition of a wiring cost minimization objective could further reduce the estimated path commute time of EDS paths. The room for reducing estimated path commute time of EDS paths exists and this gap may need to be narrowed to produce realistic CP. For instance, CP_WT, although longer than the EDS JCA path that connects 559 and 823 in terms of graph distance, has a smaller estimated path commute time (section 3.3). Finding local search paths with minimum estimated path commute time efficiently requires a clever algorithm.

EDS backtracks so there are no non-edges present in paths. However, in light of the possibility of 'jumps' between DB sites (section 3.6), jumps between nodes not connected by a PRN edge could be allowed for a local search to escape local minima. On the other hand, allowing some local search paths to be trapped in local minima may serve a purpose [26, 41]. CP do not exist for all node-pairs in KIT (section 2.5).

$C\alpha$–$C\alpha$ distance is used to drive EDS since identification of both IDS and CP rely on the motions of $C\alpha$ atoms. A reviewer commented that the use of side-chain distance has been successful in several case studies [28]. It could be worthwhile to explore the use of side-chain distance in EDS since changes in $C\beta$–$C\beta$ distance are larger than changes in $C\alpha$–$C\alpha$ distance for the key H-bond Y823-D792 (Table A3).

## 4. Conclusion

This work demonstrates that EDS paths on PRNs are biologically meaningful, and can provide biologically relevant information not readily available from BFS paths. On the KIT protein, EDS paths outperformed BFS paths on common bioinformatics tasks such as detecting pathway changes as a result of subtle conformational changes (section 3.3) and identifying mutational hotspot residues (section 3.5). The differences in EDS and BFS path properties also suggest that a local search perspective could produce better models of intra-protein communication than a global search perspective.

More importantly, this comparison initiates the compilation of a list of path properties that are characteristic of intra-protein pathways. Such a list would contribute to the design and validation of intra-protein communication models and methods in an objective and organized way. The desirable path properties identified in this work are: (i) preference for short-range over long-range edges (section 3.1); (ii) respect for the modular contours of the underlying network and reuse of sub-paths (section 3.2); (iii) sensitivity to changes in residue coordinates (section 3.3); and (iv) appropriate dependency on the integrity of protein structures purported to be responsible for long-range intra-protein communication and modulation (section 3.7).

EDS and BFS represent two fundamentally different views of proteins. The local search perspective (examined here via EDS) views proteins as navigable small-world networks and not merely networks with substantial clustering and short paths. EDS relies on the presence of not only the quantity but also the quality of clustering to produce short paths (section 3.8). While the global search view (examined here via BFS) does acknowledge the presence of clustering in proteins and uses its various forms, e.g. cliques and communities [28], to describe conformational changes associated with allosteric changes in proteins, the role clustering plays in proteins is less well integrated. A local search perspective lifts clustering beyond its hitherto topological descriptive role and demonstrates the interplay between structure and function in proteins at a deeper level in a natural way.

An elementary difference between the two perspectives lies in how clustering affects path length. Put simply, with all other things being equal, EDS paths become longer while BFS paths become shorter, with diminished clustering. This effect may be obscured by change in link density and go undetected when studying individual pathways between select pairs of residues. For this reason, we study how changes in clustering affect path length directly in the aggregate and in a controlled manner (section 3.7). Due to its adherence to the inverse relationship between clustering and path length, a local search perspective easily accommodates the notion of allosteric communication as a continuation of the protein folding process, which itself works to increase clustering in a protein while reducing inter-residue distances [9].

Finally, with its dependency on short-range connections (section 3.7) and intervention via non-hub nodes (section 3.6), our study of allosteric communication in KIT revealed its rather counter-intuitive approach to mediating and regulating communication in a network. Evidence collected from 166 PRNs further emphasized the importance of short-range edges and having the right kind of clustering distribution to navigability (section 3.8). The ability to self-form and self-regulate navigable small-world networks with short-range connections would be of practical use in artificial systems with physical or non-physical constraints on the length of connections. In an artificial system, what is 'short-range' needs to be defined or discovered though building the system. In proteins, 'short-range' is naturally defined, i.e. discovered through biological evolution. Knowing the factors that contributed to this natural definition in proteins could be insightful for artificial systems.

**Acknowledgments**
This work was made possible by the facilities of the Shared Hierarchical Academic Research Computing Network (SHARCNET:www.sharcnet.ca) and Compute/Calcul Canada. Thanks to the Dynameomics group for providing access to the MD data and to E. Laine for helpful discussions.

**Appendix A 1T45_MU and 1T45_dbMU theoretical models and their PRNs**

The theoretical models for MU and dbMU were generated with MODELLER 9.15 [42] using 1T45:A, 3G0E:A and 1PKG:A as templates for 1T45_MU, and 1T45:A and 1PKG:B as templates for 1T45_dbMU (HETATOMS were excluded). Using the PyMol align command, the Root-Mean-Squared-Differences (RMSD) between 1T45:A and the respective template proteins are estimated as follows: 0.584Å for 3G0E:A, 1.678Å for 1PKG:A and 1.581Å for 1PKG:B. The respective templates were selected, after some preliminary experimentation, for their ability to produce targets with reasonable resemblance to descriptions of MU and dbMU given in refs. [13-15]. Three evaluation criteria were considered: (i) the overall shape of their energy profiles relative to the profile for 1T45, (ii) whether the theoretical models possess the requisite secondary structures in two key areas: JM-Switch (560…570) and A-loop, and (iii) change in distance between JM-Switch and two residues in the C-lobe, 847 and 912. The target models were optimized by Modeller using the fast.refine option, but no loop modeling was applied.

For each theoretical model, five target models were generated and the one with the lowest DOPE score was selected for further analysis. 1T45_MU and 1T45_dbMU have energy profiles that are similar to the energy profile of 1T45 (Fig. A1). This provides some assurance as to the reasonableness of the target models, and is in harmony with the report in [14] that both MU and dbMU have similar overall folds to WT. The energy profile (DOPE score) is computed by MODELLER with default options. DOPE (Discrete Optimized Protein Energy) is an atomic distance-dependent statistical potential optimized for model assessment [43]. The RMSD between 1T45:A (WT) and 1T45_MU, WT and 1T45_dbMU, and 1T45_MU and 1T45_dbMU are 0.266Å, 0.176Å and 0.290Å respectively.

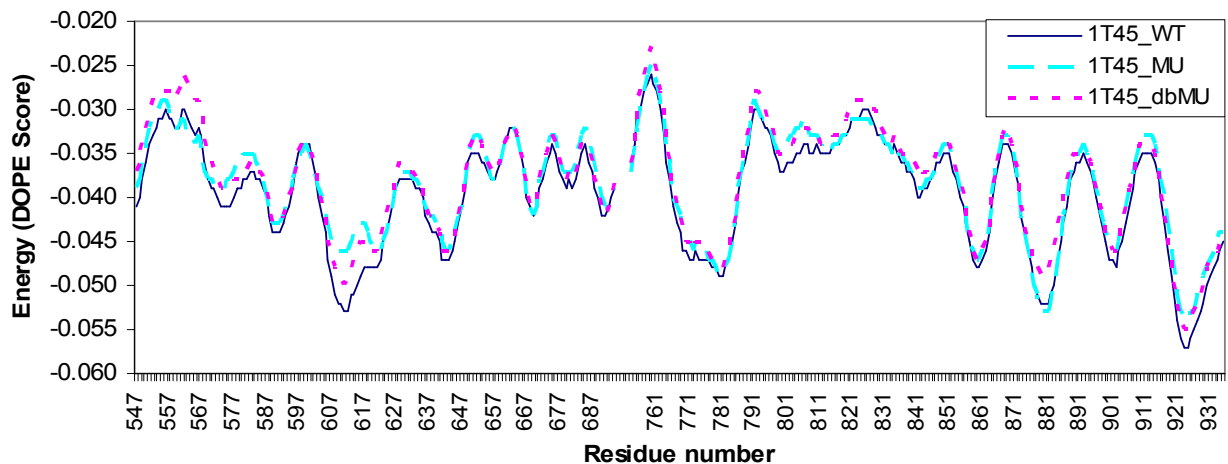

**Fig. A1** Energy profiles generated by MODELLER for WT (1T45), 1T45_MU and 1T45_dbMU.

However, both MU and dbMU possess two important structural exceptions to WT. MD simulations of MU and dbMU reveal that the small helix (817…819) in the A-loop region becomes partially unfolded in both MU and dbMU. Further, the JM-Switch region in MU exhibits an anti-parallel beta-sheet structure most of the time but this significant fold gain was not observed in the MD simulations of dbMU [14]. In agreement, 1T45_MU has more beta-residues in its JM-Switch region compared to WT, and only turn-residues at positions 817 to 819 (Table A1). 1T45_dbMU also has only turn-residues at positions 817 to 819, but its JM-Switch region is identical to WT's in terms of DSSP secondary structure assignment. The D792E mutation does not restore MU to the original WT conformation; instead dbMU is a blend of WT and MU structures [14]. Secondary structure is assigned by DSSP-2.0.4 [35, 36], and we partition the DSSP assignments into three basic groups. *Helix-residues* are residues assigned either 'H', 'I' or 'G', *beta-residues* are residues assigned either 'E' or 'B', and *turn-residues* are neither helix- nor beta-residues, i.e. encompasses coils, loops, turns and other disordered regions.

**Table A1** Number of residues by secondary structure type. WT is wild-type (inactive) KIT; 1T45_MU is our theoretical model of MU (WT activated by D816V); and 1T45_dbMU is our theoretical model of dbMU (MU deactivated by D792E).

| Secondary structure type | Overall | | | JM-Switch (560…570) | | | A-loop (810…835) | | |
|---|---|---|---|---|---|---|---|---|---|
| | WT | 1T45_MU | 1T45_dbMU | WT | 1T45_MU | 1T45_dbMU | WT | 1T45_MU | 1T45_dbMU |
| Helix (H) | 132 | 123 | 129 | 0 | 0 | 0 | 9 | 6 | 6 |
| Beta (B) | 61 | 66 | 59 | 2 | 7 | 2 | 4 | 6 | 4 |
| Turn (T) | 138 | 142 | 143 | 9 | 4 | 9 | 13 | 14 | 16 |

One of the parameters monitored in [15] by their MD simulations of MU to evidence the detachment of the JMR from the PTK as part of KIT activation is $D_{JMS}$, which is the distance between the centre-of-mass of JM-Switch and the centre-of-mass of two closest C-lobe residues (847 and 912). An increase in mean $D_{JMS}$ was observed in MU relative to WT. To obtain support for a similar phenomenon in 1T45_MU, the average distance (Cα-Cα Euclidean distance) between JM-Switch residues and residue 847 (JMS_847), and the average distance between JM-Switch residues and residue 912 (JMS_912), were measured. Both of these distances are significantly longer in 1T45_MU relative to WT (one-sided paired t.test p-values are 1.216E-06 and 1.15E-06 respectively). In WT, JMS_847 is 20.4231Å with a std. dev. of 6.1511, and JMS_912 is 29.8747Å with a std. dev. of 5.8974. In 1T45_MU, JMS_847 is 20.7342Å with a std. dev. of 6.2100, and JMS_912 is 30.2246Å with a std. dev. of 5.9935.

There are limitations to the use of theoretical models without MD simulation data. Proteins, even when in their "native state", are dynamic spatial networks. So both the cystrallized structure of 1T45 and the theoretical models of MU and dbMU are at best merely theoretical snapshots of the respective protein's behavior. We do not claim that the theoretical models or their respective PRNs capture all the salient details of WT, MU or dbMU. For instance, the unfolding of the small helix (817…819) in MU is further evidenced by an increase in Cα-Cα Euclidean distance between 816 and 823 relative to WT [15]. However, this distance decreased in 1T45_MU relative to WT. The Cα-Cα Euclidean distance between 816 and 823 is 10.1687Å in WT, 9.90216Å in 1T45_MU and 10.089Å 1T45_dbMU. To reiterate, the main objective of this work is to compare EDS with BFS paths on PRNs; not to reproduce results from [13-15] in all of its detail or to gain new biological insights into the functioning of KIT, although this would be nice. The KIT allostery studies cited here provide a biological context for this comparison.

A PRN is built for each KIT variant: WT (1T45), 1T45_MU and 1T45_dbMU. These three PRNs are examined to evaluate how well they manifest the KIT mutants by comparing the Cα-Cα Euclidean distance of several PRN edges that correspond to the H-bonds mentioned in [14] and [15]. The following observations serve to reassure that the PRNs are reasonable and meaningful in the main, with respect to WT, MU and dbMU.

In its autoinhibited inactive form (WT), the JMR is a mainly solvent-exposed coil, attached to the C-lobe of the PTK by four stable H-bonds: V560-N787, K558-I789, Y568-F848 and Y570-Y846 [15]. Post-phosphorylation of its tyrosine residues, the JMR is likely to be less attached to the PTK [44], and in MU (D816V), the JMR becomes more ordered with its JM-Switch region taking on a well-formed anti-parallel beta-sheet structure [13]. In contrast, the JM-Switch region of dbMU (D816V/D792E) is less ordered and resembles the JM-Switch of WT [14]. We confirm that these four residue pairs are linked in all three PRNs. Three of these four edges are longer in 1T45_MU than in WT, but are shorter in 1T45_dbMU than in 1T45_MU (Table A2). These distance changes suggest the JMR has displaced itself relative to the PTK in the expected direction. The formation of the beta-sheet in the JM-Switch region is evidenced via two intra-JMR H-bonds: V559-V570 and E561-V569, which are rarely observed in the MD simulation of WT, but are highly present (stable) in the MD simulation of the KIT mutants [15]. These two residue pairs are also linked in all three PRNs, and changes in their distances concur with observations from the MD simulations. Both edges are shorter in 1T45_MU relative to WT, and longer in 1T45_dbMU relative to 1T45_MU (Table A2).

**Table A2** Cα-Cα Euclidean distance (units in Angstrom Å) of PRN edges associated with H-bonds that are related to changes in the JMR [15].

|  | PRN edge | WT | 1T45_MU | 1T45_MU – WT | 1T45_dbMU | 1T45_dbMU – 1T45_MU |
|---|---|---|---|---|---|---|
| Distance increases evidence the JMR is less attached to the PTK in 1T45_MU. | 560-787 | 5.63535 | 5.56056 | -0.07479 | 5.76068 | 0.20012 |
|  | 558-789 | 5.29895 | 5.35059 | 0.05164 | 5.12207 | -0.22852 |
|  | 568-848 | 9.11620 | 9.69018 | 0.57398 | 9.2965 | -0.39368 |
|  | 570-846 | 9.84692 | 10.4424 | 0.59548 | 9.88363 | -0.55877 |
| Distance decreases evidence the beta-sheet formation in the JM-Switch region of 1T45_MU. | 559-570 | 5.77561 | 5.71981 | -0.0558 | 5.74989 | 0.03008 |
|  | 561-569 | 5.64005 | 5.40700 | -0.23305 | 5.66234 | 0.25534 |

Three very stable H-bonds were identified in [14] as playing a key role in the allosteric communication between the JMR and A-loop regions of inactive KIT: Y823-D792, D792-H790, and H790-N797. The stability (occupancy rates) for these three H-bonds in the MD simulation of WT were 95%, 95%, and 93% respectively. The D816V mutation affects the stability of these three H-bonds to different extents. In the MD simulation of MU, the stability of Y823-D792 deteriorated to 45%, while N797 showed an increased preference for interacting with D792 (85%) over H790. The stability of Y823-D792 is restored in dbMU, but N797 spends almost equal time bonding with H790 and D792. Hence, dbMU's H-bond network between Y823 and the catalytic-loop residues (H790, D792 and N797) is a hybrid of the same H-bond network in WT and in MU [14]. H790-D792 is a very stable bond in all conformational ensembles of the three KIT variants. We confirm that all four residue pairs mentioned here are linked in all three PRNs, and present evidence for these local effects of the point mutations in terms of changes in Euclidean distance (Table A3). The Euclidean distance between 823 and 792 increased in 1T45_MU (relative to WT) but decreased in 1T45_dbMU (relative to 1T45_MU). The Euclidean distance between 790 and 797 increased while the Euclidean distance between 792 and 797 decreased in 1T45_MU. The Euclidean distance of both 792-797 and 790-797 decreased in 1T45_dbMU.

**Table A3** Increases (decreases) in Cα-Cα and Cβ–Cβ Euclidean distances (units in Angstrom Å) evidence the deterioration (improvement) in stability of select H-bonds reported in [14].

| PRN edge | WT | | 1T45_MU | | 1T45_dbMU | |
|---|---|---|---|---|---|---|
|  | Cα-Cα | Cβ–Cβ | Cα-Cα | Cβ–Cβ | Cα-Cα | Cβ–Cβ |
| 792-823 | 10.3172 | 9.1561 | 10.4965 | 10.2520 | 10.4924 | 9.7760 |
| 790-792 | 5.5726 | 6.1617 | 5.4655 | 5.7265 | 5.4349 | 5.7161 |
| 790-797 | 9.7923 | 7.5358 | 9.9501 | 7.5433 | 9.8715 | 7.4993 |
| 792-797 | 7.4072 | 5.7035 | 7.3048 | 5.2225 | 7.2352 | 5.1501 |

| PRN edge | 1T45_MU – WT | | | 1T45_dbMU – 1T45_MU | | |
|---|---|---|---|---|---|---|
|  | Cα-Cα | Cβ–Cβ | Total | Cα-Cα | Cβ–Cβ | Total |
| 823-792 | 0.1793 | 1.0959 | 1.2752 | -0.0041 | -0.4760 | -0.4801 |
| 792-790 | -0.1071 | -0.4352 | -0.5423 | -0.0306 | -0.0104 | -0.0410 |
| 790-797 | 0.1578 | 0.0075 | 0.1653 | -0.0786 | -0.0440 | -0.1226 |
| 792-797 | -0.1024 | -0.4810 | -0.5834 | -0.0696 | -0.0724 | -0.1420 |

To support our use of Cα-Cα and Cβ–Cβ Euclidean distance as proxy for interaction stability in an MD simulation, we use the 2EZN dataset from section 3.1 to obtain the following correlations: the Cα-Cα and Cβ–Cβ Euclidean distance of PRN0 edges are significantly and negatively (-0.5118428 and -0.4924738 respectively) correlated with edge stability. The average Cα-Cα Euclidean distance of PRN0 edges with stability > 0.5 is 7.770Å (std. dev. 1.949029), while for edges with stability ≤ 0.5, it is 10.150Å (std. dev. 1.670488). The average Cβ –Cβ Euclidean distance of PRN0 edges with stability > 0.5 is 7.969Å (std. dev. 1.902419), while for edges with stability ≤ 0.5, it is 10.220Å (std. dev. 1.43093). Hence shorter PRN0 edges tend to be more stable and vice versa. The edges of a PRN0 tend to be stable in general. 83.63% (700/837) of 2EZN PRN0 edges have stability > 0.5.

**Appendix B Pseudo-code for EDS**

**EDS**(*s*, *t*)
Input: source node *s*, target node *t*, a graph *G*
Output: EDS path *p*     //A sequence of nodes in order of EDS visit with *s* as the leftmost node and *t* as
                //the rightmost node when EDS is successful.
Main variable: *inspected*     //The set of direct neighbors of nodes currently in *p*, excluding the nodes in
                //*p*, sorted in ascending order by their Euclidean distance to *t*. The leftmost
                //node of *inspected* is a node currently closest to *t* and not already in *p*.

```
 1:  append s to p    //p = ⟨s⟩
 2:  inspected is empty initially
 3:  do
 4:     x := the rightmost node of p
 5:     for each node i in the set of direct neighbors of x in G do
 6:        if i = t then
 7:           append i to p
 8:           stop     //path p from s to t is found
 9:        end if
10:        if i is not in p then
11:           calculate distance d between i and t
12:           add i to inspected   //nodes in inspected are sorted by their distance to t
13:        end if
14:     end for
15:     let y be the leftmost node in inspected   //closest to t amongst all nodes currently in inspected
16:     if x and y are not edgeed in G then
17:        inspect p from right to left starting at x for a node z that is a direct neighbor of y   //backtrack
18:        append to p the sub-path of p starting from x to z   //p = ⟨s…z…x…z⟩
19:     end if
20:     append y to p      //p = ⟨s…z…x…zy⟩
21:     remove y from inspected
22:  while inspected is not empty
23:  //path p from s to t is not found
```

**Appendix C Path stability and path commute time**

Snapshots of the native dynamics (298K) of 12 randomly selected proteins were downloaded from the Dynameomics database [24, 25]. A PRN0 is constructed for the chain of each protein within the residue range simulated in Dynameomics (Table C1). Except for 2EZN where the entire MD simulation is used, stability and commute times of edges are computed using the first $x$ of the $y$ available MD native dynamics snapshots (this is due to data download constraints). With these edge stability and edge commute time information, we evaluated the path stability and path commute time for the set of all EDS and BFS paths generated by each PRN0.

**Table C1** PDB code and basic statistics for the 12 proteins.

| PDB code (residue range used) | MD snapshots used/total | Number of PRN0 nodes | Number of PRN0 edges | | Number of paths with > 1 edge in PRN0 | |
|---|---|---|---|---|---|---|
| | | | Short-range (*SE*) | Long-range (*LE*) | Short-range (*SP*) | Long-range (*LP*) |
| 1CUK-A (156-203) | 20,000/51,163 | 48 | 178 | 65 | 494 | 1,276 |
| 1EZG-A (2-83) | 20,000/52,490 | 82 | 326 | 324 | 878 | 4,464 |
| 1ELP-A (1-83) | 20,000/52,318 | 83 | 215 | 328 | 1,120 | 4,600 |
| 2EZN-A (1-101) | 51,000/51,000 | 101 | 346 | 491 | 1,218 | 7,208 |
| 3GRS-A (366-478) | 20,000/53,650 | 113 | 330 | 329 | 1,490 | 9,848 |
| 1EBD-A (155-271) | 20,000/53,224 | 117 | 335 | 354 | 1,560 | 10,634 |
| 1D0N-A (27-159) | 20,000/51,311 | 133 | 386 | 431 | 1,778 | 14,144 |
| 1IHB-A (5-160) | 20,000/51,867 | 156 | 587 | 489 | 1,836 | 20,192 |
| 1BFD-A (2-181) | 20,000/52,997 | 180 | 561 | 712 | 2,368 | 27,306 |
| 1ESJ-A (1-272) | 20,000/52,274 | 272 | 939 | 1,065 | 3,452 | 66,252 |
| 1BS2-A (136-482) | 12,000/51,989 | 347 | 1,208 | 1,164 | 4,414 | 110,904 |
| 1EHE-A (5-404) | 12,000/51,560 | 399 | 1,430 | 1,391 | 5,010 | 148,150 |

**Table C2 p-values generated with R's Wilcoxon one-sided test, paired when possible (path comparisons).** For all the 12 PRN0s, *LP* EDS paths are significantly (p-value < 0.05) more stable than *LP* BFS paths, and *LP* EDS paths have significantly smaller path commute time than *LP* BFS paths. Except for 1EZG, *SE* are significantly more stable and have significantly smaller commute time than *LE*.

| PRN0 | *LP* path stability | *LP* path commute time | Edge stability | Edge commute time |
|---|---|---|---|---|
| | BFS < EDS | BFS > EDS | *SE* > *LE* | *SE* < *LE* |
| 1CUK-A | 2.07E-02 | 5.20E-19 | 2.68E-09 | 1.32E-25 |
| 1EZG-A | 9.02E-29 | 1.45E-02 | 3.68E-01 | 2.57E-07 |
| 1ELP-A | 2.13E-09 | 4.21E-25 | 1.83E-03 | 1.39E-09 |
| 2EZN-A | 9.45E-23 | 2.53E-82 | 8.40E-31 | 1.80E-52 |
| 3GRS-A | 4.42E-51 | 1.90E-106 | 4.94E-14 | 1.87E-24 |
| 1EBD-A | 5.35E-60 | 3.57E-115 | 3.31E-10 | 9.06E-31 |
| 1D0N-A | 5.25E-46 | 7.34E-86 | 3.05E-19 | 9.84E-29 |
| 1IHB-A | 1.18E-59 | 2.11E-140 | 1.46E-73 | 2.40E-103 |
| 1BFD-A | 1.30E-25 | 4.61E-236 | 2.68E-43 | 5.09E-63 |
| 1ESJ-A | 1.57E-165 | 2.22E-267 | 8.06E-67 | 9.46E-147 |
| 1BS2-A | 2.33E-281 | 0.00E+00 | 1.01E-113 | 5.28E-187 |
| 1EHE-A | 0.00E+00 | 0.00E+00 | 4.92E-143 | 8.34E-247 |

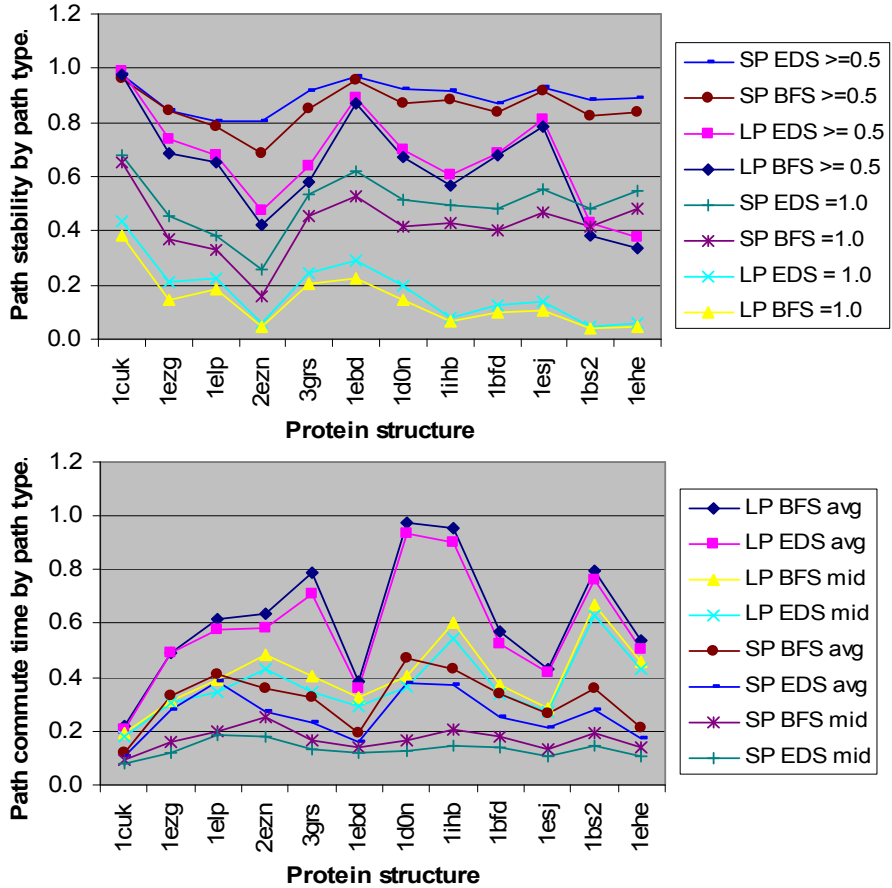

*SP EDS >= 0.5* denotes the fraction of EDS short-range paths with > 1 edge with path stability of at least 0.5.

*LP BFS = 1.0* denotes the fraction of BFS long-range paths with > 1 edge with path stability of 1.0.

*LP BFS avg* denotes the average path commute time of long-range BFS paths.

*SP EDS mid* denotes the median path commute time of short-range EDS paths.

**Fig. C1 Top:** Regardless of search type (BFS or EDS), short-range paths are more stable than long-range paths. Regardless of path range (short or long), EDS paths are more stable than BFS paths. **Bottom:** Regardless of search type (EDS or BFS), long-range paths have longer commute times than short-range paths. Regardless of path range (short or long), BFS paths have longer commute times than EDS paths.

**Appendix D Compressibility of paths/walks and reducibility of path sets**

A random walk (*RW*) on a network with modular structure will spend more time wandering amongst nodes of a module than traversing between modules [45]. This is the common principle exploited by flow-based clustering algorithms. A RW moves to a direct neighbor of the current node, selected uniformly at random, until the target node is found. It is expected then that if each module in a network were labeled uniquely but nodes of the same module were labeled identically, and a path *p* is a sequence of node labels in the order the nodes are visited by *p*, paths which follow the modular contours of a network more faithfully would be more *compressible*. A path represented as a string of symbols is compressible if it has a sub-string of length greater than one that is comprised of identical symbols. Let nodes(*p*) be the number of nodes in path *p*. The *compression ratio* for a path is $cr(p) = [nodes(p) - nodes(cp)] / nodes(p)$ when nodes(*cp*) > 1, and $cr(p) = 1$ when nodes(*cp*) = 1. Larger *cr* values imply greater compression. *cr* = 0 when there is no compression, and a path with maximum compression (*cr* = 1) stays within a single module.

A set of paths has better reducibility if a larger proportion of its paths are sub-paths of other paths in the set. A path that is not a sub-path of any other path is a non-reducible path. Let |*P*| be the number of unique paths in a path set *P*, and *q* the number of non-reducible paths in *P*, then the *reducibility ratio* (*rr*) of *P* is |*P*|/*q*.

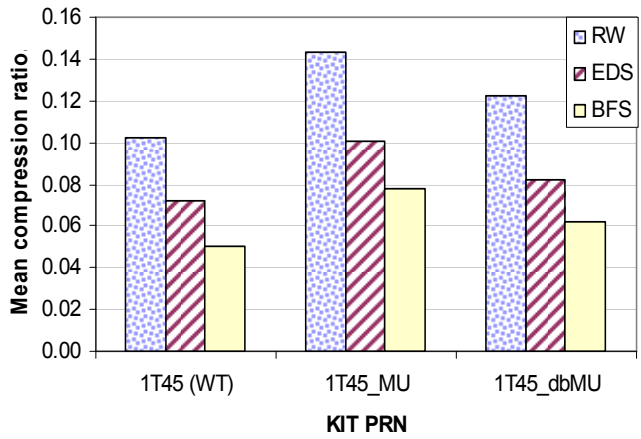

**Fig. D1** Compressibility of random walks (RW) evidence the presence of modules in terms of IDSs. EDS paths are significantly more compressible than BFS paths.

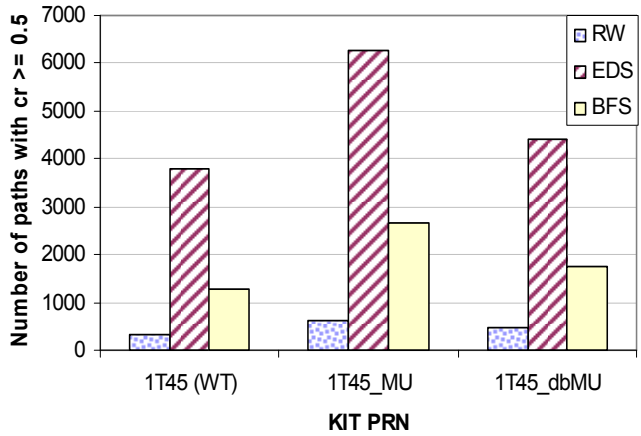

**Fig. D2** EDS produces at least twice as many highly ($cr \geq 0.5$) compressed paths than BFS, and about 10 times more highly compressed paths than RW.

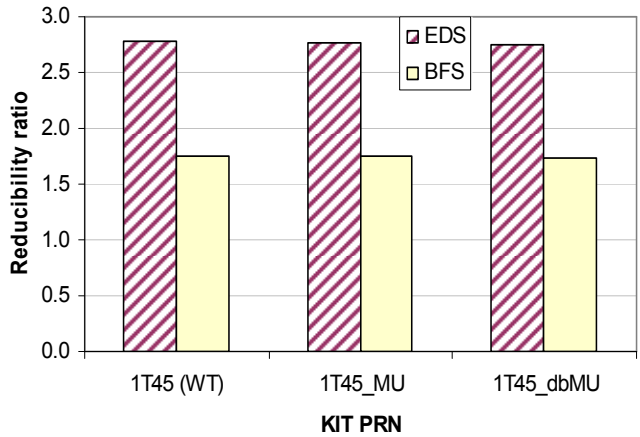

**Fig. D3** EDS paths are more reducible than BFS paths.

**Appendix E closeness base on estimated path commute time**

$Closeness(x^i) = \sum_{y \neq x}^{N}[epct(x^i, y^j) + epct(y^j, x^i)]$ gives the closeness of node $x$ in region $i$ to all other nodes in region $j$. In Fig. E1, $x$ and $y$ are restricted to two regions: the N-lobe (582…684) and the C-lobe (762…935).

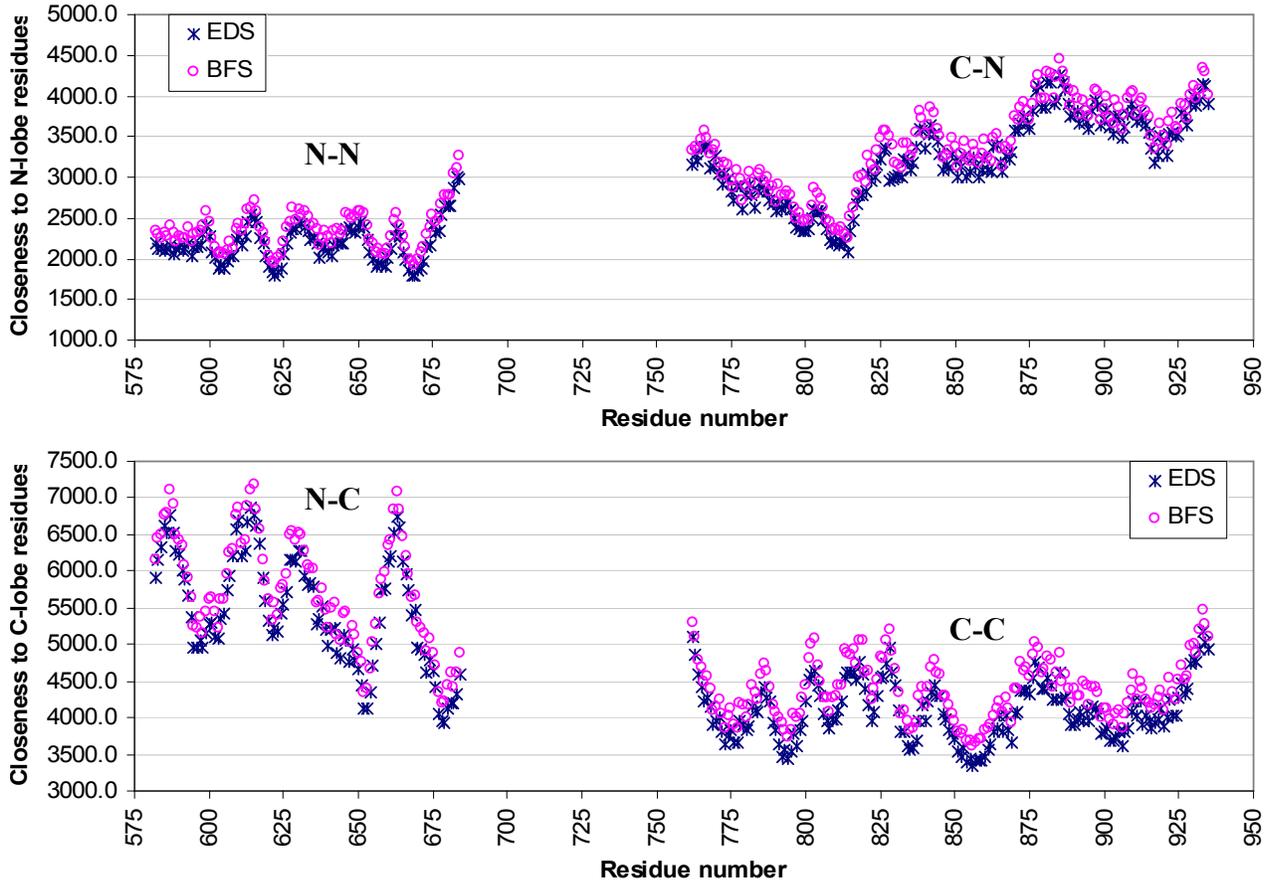

**Fig. E1** Closeness of residues within and between the two lobes of WT'S PTK.

**Appendix F**

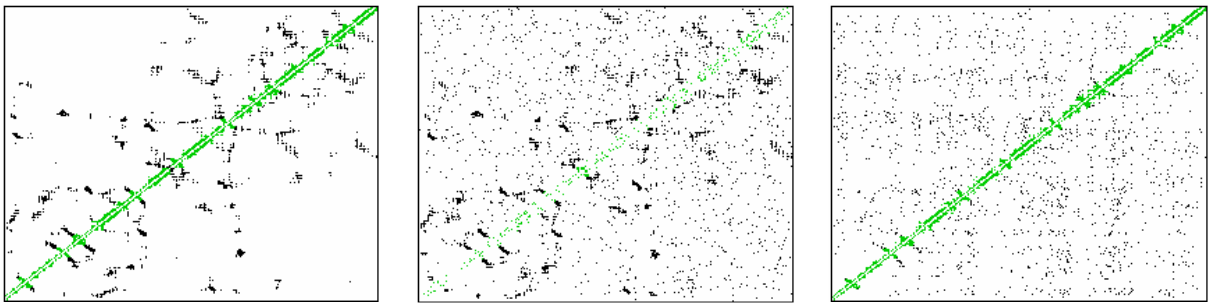

**Fig. F1** From left to right: Contact map (adjacency matrix) of the 1T45 (WT) PRN, a randSE network and a randLE network. A non-white cell (x, y) denotes the presence of an edge (x, y). Short-range edges are marked in green.

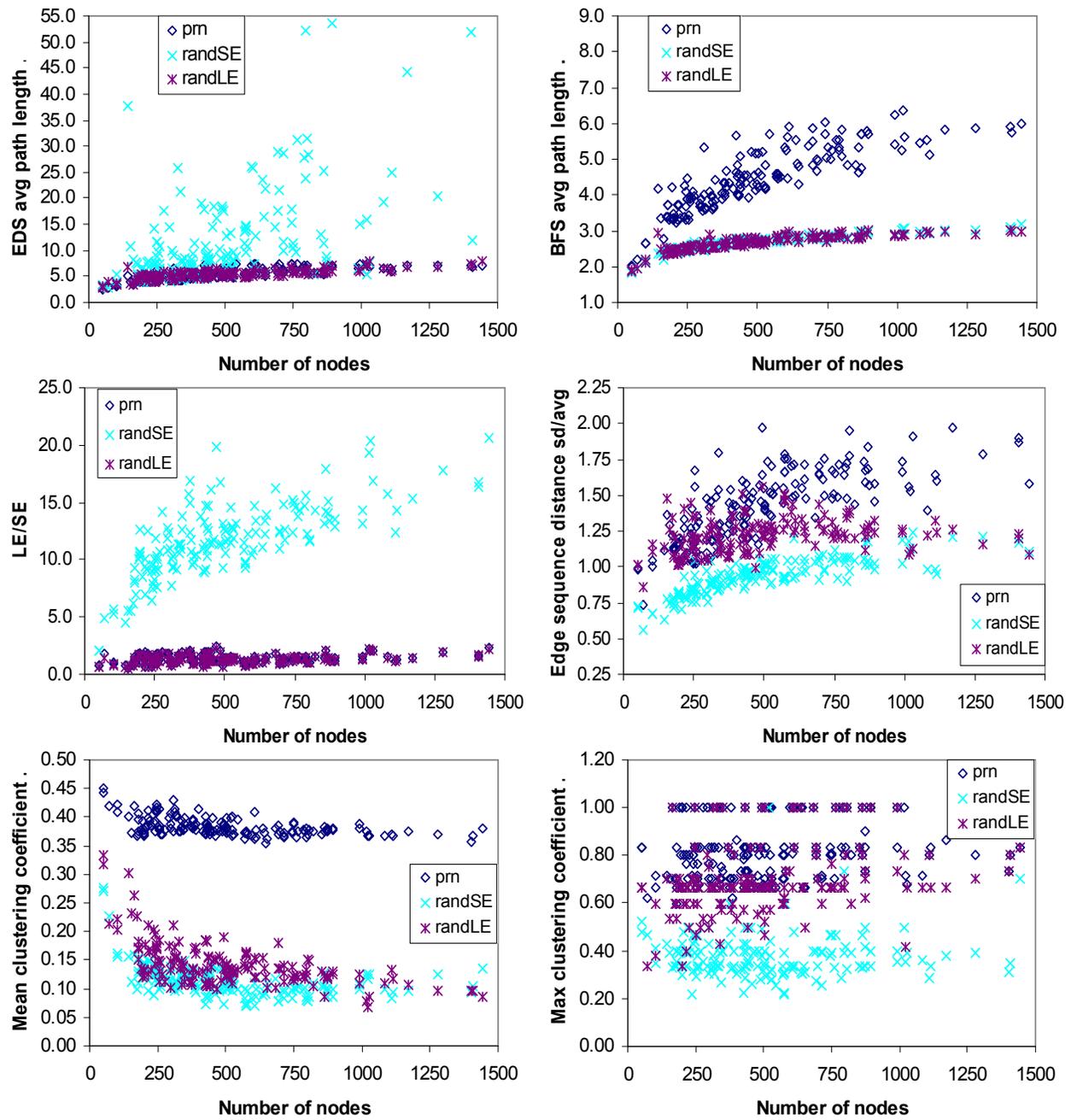

**Fig. F2** Network statistics for the 166 PRNs in [9] and their randSE and randLE networks.

**Appendix G Residue interaction networks (RIN)**

A RIN is a network representation of a protein in which a pair of nodes (each representing an amino acid) is connected by an edge if the Euclidean distance between its endpoint Cα atoms is within a user specified threshold range. In [9], we showed that PRNs are weaker expander graphs than RINs. Here, we elaborate on this point in a more concrete manner. To be compatible with PRNs, peptide bonds are excluded from RINs.

Several of the edges in Tables A2 and A3 have rather large Cα-Cα distance values. Using larger distance cut-offs can be problematic for RINs as important structural variations may be smeared away. We illustrate this problem on 1T45 with the WT PRN and two WT RINs: one with a 10.3172Å distance cutoff (just long enough for 792-823 to be an edge), and another with a cutoff distance of 15.5677Å (which is the maximum Cα-Cα Euclidean distance of edges in the WT PRN). Fig. G1 illustrates that the RINs have many more edges that cross large cavities on the molecular surface. In contrast, the PRNs are less susceptible to this problem.

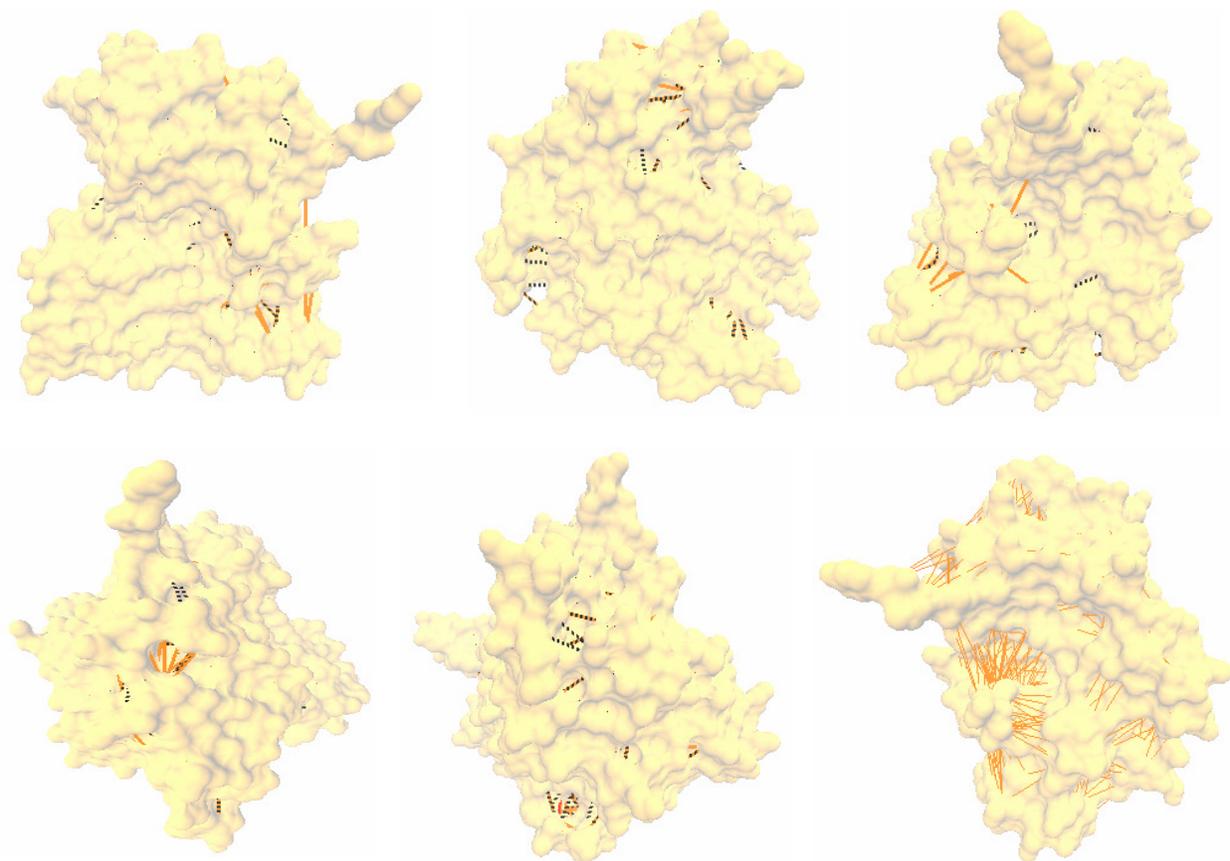

**Fig. G1** Six views of 1T45 showing cavity crossings by PRN and RIN edges. The surface of the 1T45 molecule was drawn with VMD 1.9.2 [46] SURF at a probe radius of 1.4 Å. The dotted lines in black are PRN edges. The orange solid lines are RIN edges. The 1T45 RIN edges in the first five panels have a maximum Euclidean distance (measured Cα to Cα) of 10.3172Å (just long enough for 792-823 to be an edge). This RIN has 2906 edges. The 1T45 PRN edges have a maximum Euclidean distance (measured Cα to Cα) of 15.5677Å. The bottom rightmost panel is a view of 1T45 with RIN edges at a cutoff of 15.5677 Å. This RIN is more dense, it has 9139 edges and show many more cavity crossing edges.

When the analysis in section 3.6 is repeated on WT's RIN (10.3172Å cutoff distance), the same conclusions about network structural differences between IDS and hub nodes can be made (Table G1). However, there is significantly less distinction between the secondary structures in WT's RIN even though it has a smaller maximum edge distance than WT's PRN (Table G2). In WT's PRN, both helix- and beta-residues have significantly smaller clustering coefficients on average than the turn-residues (p-values = 3.431e-05, 1.081e-06). But at $\alpha = 0.05$, both helix- and beta- residues in WT's RIN do not have significantly smaller clustering coefficients than the turn-residues (p-values = 0.07039, 0.1861). This reinforces the point about structural blurring in RINs. The WT RIN (2906) has more edges than the WT PRN (2314), but clustering and modularity also depends on the placement of edges in the network.

**Table G1** Node degree and node clustering coefficient (mean ± std. dev.) for WT residues by node type.

| Node Type | Number of nodes | WT PRN | | WT RIN 10.3172Å distance cutoff | |
|---|---|---|---|---|---|
| | | Node degree | Node clustering coefficient | Node degree | Node clustering coefficient |
| All | 331 | 13.98 ± 5.2766 | 0.3900 ± 0.1019 | 17.56 ± 6.5560 | 0.4197 ± 0.0999 |
| Hubs | 71 | 18.01 ± 4.3408 | 0.3568 ± 0.0483 | 23.0 ± 5.8089 | 0.3828 ± 0.0679 |
| IDS | 101 | 9.743 ± 4.1028 | 0.4402 ± 0.1388 | 11.57 ± 4.4505 | 0.4598 ± 0.1329 |

**Table G2** Node degree and node clustering coefficient (mean ± std. dev.) for WT residues by secondary structure.

| Node type | Number of nodes | WT PRN | | WT RIN 10.3172Å distance cutoff | |
|---|---|---|---|---|---|
| | | Node degree | Node clustering coefficient | Node degree | Node clustering coefficient |
| Helix | 132 | 15.35 ± 5.5327 | 0.3701 ± 0.0794 | 18.88 ± 6.4524 | 0.4109 ± 0.0801 |
| Beta | 61 | 15.36 ± 3.3468 | 0.3606 ± 0.0503 | 20.93 ± 5.0096 | 0.4167 ± 0.0757 |
| Turn | 138 | 12.07 ± 5.1520 | 0.4220 ± 0.1268 | 14.80 ± 6.1914 | 0.4295 ± 0.1233 |

## Appendix H

**Table H1** List of residues in each IDS and the region in which each IDS is situated [14]. # is number of residues in an IDS.

| Overlapping region | IDS | WT | # | MU (D816V) | # | dbMU (D816V, D792E) | # |
|---|---|---|---|---|---|---|---|
| JM-Proximal | S1 | 547…554 | 8 | 547…553 | 7 | 547…552 | 6 |
| JM-Switch | S2 | 561…569 | 9 | 561…570 | 10 | 561…569 | 9 |
| JM-Zipper | S3 | 574…581 | 8 | 571…577 | 7 | 572…581 | 10 |
| Loop-I | S4 | 588, 609…618 | 11 | 588, 609…618 | 11 | 588, 608…618 | 12 |
| C-loop-I C-helix (631…647) | S5 | 626…633 | 8 | 598…601, 625…635 | 15 | 626…631 | 6 |
| Loop-II | S6 | 585…587, 661…666 | 9 | 586, 587, 661…666 | 8 | 586, 587, 661…665 | 7 |
| Pseudo-KID | S7 | 688…694, 753…762 | 17 | 684, 687, 690…693, 753…763 | 17 | 687…694, 753…762 | 18 |
| A-loop (810…835) | S8 | 824…831 | 8 | 816…832 | 17 | 814…821, 824…831 | 16 |
| G-helix (865…885) (binding site) | S9 | 870…882 | 13 | 870…883 | 14 | 874…887 | 14 |
| C-tail (930…935) | S10 | 926…935 | 10 | 926…935 | 10 | 925…935 | 11 |
| Total number of IDS residues | | | 101 | | 116 | | 109 |